\documentclass[11pt]{article}

\usepackage[margin=.75in]{geometry}
\usepackage{multirow}
\usepackage{datetime}
\usepackage[dvipsnames]{xcolor}
\usepackage{amsmath, amsthm, amssymb,fancyhdr,mathrsfs, slashbox, enumerate}
\usepackage{latexsym}
\usepackage[round, authoryear]{natbib}
\usepackage{hyperref}
\usepackage{MnSymbol}
\usepackage{graphicx}
\graphicspath{ {./images/} }
\title{Figures}

\renewcommand{\P}{\mathbb{P}}
\newcommand{\balpha}{\boldsymbol{\alpha}}
\newcommand{\bxi}{\boldsymbol\xi}
\newcommand{\bpi}{\boldsymbol\pi}
\newcommand{\bt}{{\bf\theta}}
\newcommand{\bpsi}{{\bf\Psi}}
\newcommand{\bd}{{\bf \Delta}}
\newcommand{\dd}{{\boldsymbol \delta}}
\newcommand{\ld}{\Delta}
\newcommand{\ldd}{\delta}
\newcommand{\bomega}{{\boldsymbol\Omega}}

\newcommand{\I}{\mathbb{I}}

\begin{document}


\begin{center}
      \Large{\bf\sc Bayesian Mixture Modelling with Ranked Set Samples}
\end{center}

\begin{center}
  \noindent{          {\sc Amirhossein Alvandi}$^{\dagger}$,
  {\sc Sedigheh Omidvar}$^{\ddagger}$,
  {\sc Armin Hatefi}$^{ \diamond,}$\footnote{\footnotesize{ Corresponding
            author.\\  E-mail: ahatefi@mun.ca, Phone: +1 (709) 864-8416.}}},
            {\sc Mohammad Jafari Jozani}$^{\star}$,
              {\sc Omer Ozturk}$^{\Im}$
              and  
              {\sc Nader Nematollahi}$^\ddagger$

\end{center}

\noindent{ \footnotesize{\it $\dagger$ Department of Mathematics and Statistics, University of Massachusetts, Amherst, MA, USA. }}\\
\noindent{ \footnotesize{\it $\ddagger$ Department of Statistics, Allameh Tabataba'i University, Tehran, Iran. }}\\
\noindent{ \footnotesize{\it $\diamond$ Department of Mathematics and Statistics, Memorial University of Newfoundland,
                        St. John's, NL, Canada. }}\\
\noindent{\footnotesize{\it $\star$ Department of Statistics,  University of Manitoba,
                        Winnipeg, MB, Canada. }}\\
\noindent{\footnotesize{\it $\Im$ Department of Statistics, The Ohio State University,
                            1958 Neil Avenue, Columbus, OH, USA. }}                            

\begin{center} {\small \bf Abstract}:  \end{center}

\footnotesize{We consider the Bayesian estimation of the parameters of a finite mixture model from independent order statistics arising from imperfect ranked set sampling designs. As a cost-effective method, ranked set sampling enables us to incorporate easily attainable characteristics, as ranking information, into data collection and Bayesian estimation. To handle the special structure of the ranked set samples, we develop a Bayesian estimation approach exploiting the Expectation-Maximization (EM) algorithm in estimating the ranking parameters and Metropolis within Gibbs Sampling to estimate the parameters of the underlying mixture model. Our findings show that the proposed RSS-based Bayesian estimation method outperforms the commonly used Bayesian counterpart using simple random sampling. The developed method is finally applied to estimate the bone disorder status of women aged 50 and older. }



\normalsize

\section{ Introduction } \label{sec:intro}
Osteoporosis is a major health problem characterized by a significant reduction in mass 
and microarchitecture of bone tissues \citep{center1999mortality,consensus2001development}. 
The disease increases the propensity to skeletal 
fragility and osteoporotic fractures in various body areas like the femur, hip and spine \citep{melton1997epidemiology}. 
Osteoporosis inflicts significant medical and socioeconomic costs on health systems  
\citep{burge2007incidence,melton1992perspective,haussler2007epidemiology}. 
It affects the life quality of patients such that 40\% of patients 
with osteoporosis develop lifetime fractures, most commonly in the hip and spine. Up to 
$\%20$ of patients with osteoporosis fracture in the spine or hip end up with 12-month 
excess mortality from subsequent medical complications such as pneumonia due to chronic immobilization \citep{center1999mortality}. 
In the case of economic burden,  it is reported, for example, the direct annual cost of 
osteoporosis-related fractures accounts for, on average, between \$5000 to \$6500 billion 
in Canada, the USA and Europe, excluding the indirect costs such as their related disabilities \citep{pike2011economic,viswanathan2012direct,hopkins2016current}. 
On the other hand, as a silent thief, osteoporosis occurs insidiously. For instance, three 
out of four South Korean women are unaware of their osteoporosis. Osteoporosis is typically 
diagnosed after the first clinical fracture has occurred \citep{vestergaard2005osteoporosis}. 
 According to the advancing aging 
population, it is critical to study osteoporosis and plan careful measures to monitor the 
well-being and life quality of the aged groups in the community.    
 
According to WHO and Osteoporosis National Foundation (ONF), bone mineral density (BMD)
 is considered one of the most valid and reliable methods to diagnose osteoporosis 
 status. BMD measurements, given by T-scores, are calculated by the dual-energy 
 X-ray absorptiometry (DXA), which is a costly and time-consuming procedure 
 \citep{kanis2002diagnosis,who2003prevention}. Despite the challenge, practitioners 
 typically have access to various  easy-to-measure
 characteristics about the patients such as weight, age, BMI, and BMD scores from 
 previous years \citep{de2005body,cummings1995risk}. Although plenty of individuals 
 are susceptible to osteoporosis in the community, osteoporosis studies are sometimes 
 limited to analysis based on relatively small sample sizes due to the cost of BMD measurement. 
Ranked set sampling (RSS), as a cost-effective method, can be employed as a remedy to 
obtain more informative samples from the osteoporosis population. RSS enables us to 
incorporate information contained in the in-expensive characteristics (e.g., patient age) 
as ranking information into data collection from the BMD population. This can augment the 
sample of small size from the BMD population and consequently leads to more efficient 
estimate of the population characteristics.   
 Unlike simple random sampling (SRS), RSS design creates artificial ranking 
strata over the population and allows to draw samples from all aspects of the population. 
RSS has found applications in various research areas such as nonparametric statistics \citep{frey2012nonparametric,ozturk2013combining,zamanzade2017estimation}, clustered randomized designs \citep{wang2016using,ozturk2023models},  breast cancer \citep{hatefi2017improved}, 
behavioural science \citep{helu2011nonparametric} and fishery \citep{hate2020finite} 
to name a few. For more detail about the theory and applications, readers are 
referred to \cite{chen2013ranked}.

Finite mixture models (FMMs), as a powerful and flexible tool, play a crucial role 
in classifying and analyzing heterogeneous populations \citep{peel2000finite}. FMMs 
have typically arisen from commonly used simple random sampling. In many medical 
applications like osteoporosis research, measuring response variables (e.g., the 
status of disease) is costly; however, various inexpensive laboratory or demographic 
characteristics are available that are associated with the response variable. In these 
situations, analyzing FMMs under a more informative sampling design, such as ranked set 
sampling, is more desirable. 
Accordingly, \cite{hatefi2014estimation} developed a parametric inference of FMMs based 
on RSS data assuming no error is involved in data collection. \cite{hatefi2013fisher} computed the 
fisher information travelling between complete and incomplete FMMs under perfect and imperfect RSS designs. \cite{hatefi2015mixture} and \cite{omidvar2018judgment} studied the FMMs from partially ranked set 
samples and judgmental post-stratification samples, respectively. 

This manuscript explores estimating the parameters of a finite mixture of normal 
distributions from imperfect RSS in a Bayesian framework. The Gibbs sampling is 
one of the most common approaches 
in estimating the parameters of mixture models \citep{diebolt1994estimation,richardson1997bayesian,chib1995marginal}. 
The Gibbs sampling promotes the data augmentation \citep{tanner1987calculation} 
by allowing us to sample 
iteratively from univariate conditional distributions of the parameters, ranking 
strata and component membership latent variables rather than the joint distributions. 
Unlike the  Gibbs sampler based on SRS data, 
the posterior distribution of the component parameters of the mixture population can 
not be evaluated directly under the RSS-based Gibbs sampling. To deal with this problem, 
we employ the Metropolis-Hastings approach \citep{tierney1994markov,chib1995understanding,robert1999monte} 
to accept-reject candidates from the target posterior distributions of the component parameters 
within each iteration of the RSS-based Gibbs sampler. Through extensive numerical studies, 
we compare the performance of the SRS-based Gibbs sampling and RSS-based Metropolis-within-Gibbs 
sampling in estimating the parameters of a mixture of normal distributions. We observe that 
the Bayesian RSS estimators outperform their SRS counterparts in estimating the parameters 
of the mixture population. We finally apply the  Bayesian estimators to analyze the bone 
mineral data of women aged 50 and older.

This manuscript is organized as follows. Section \ref{like} introduces the missing mechanisms 
and incomplete and complete likelihood functions of FMMs under SRS and RSS data. Section 
\ref{post} develops the SRS-based Gibbs sampling and RSS-based Metropolis-within-Gibbs sampling
 for a finite mixture of normal distributions. Section \ref{sim} evaluates the performance of 
 the Bayesian RSS and SRS estimators through simulation studies. We applied the developed 
 Bayesian methods to analyze bone mineral data in Section \ref{bone}. Finally, we present 
 the summary and concluding remarks in Section \ref{sum}.

 \section{Likelihood functions from RSS data}\label{like}
Let $X$ denote a random variable of interest describing the random phenomenon of interest. 
Suppose that the probability density function (pdf) of $X$ follows  a finite mixture model (FMM) consisting of  $J$  components with the vector of mixing proportions 
$\boldsymbol{\pi}=(\pi_1, \ldots, \pi_J)$, with $\pi_j>0$,
$\sum_{j=1}^J \pi_j=1$ is given by 
\begin{eqnarray}\label{mixture}
      f(x; {\bf \Psi})= \pi_1 f_1(x ; {\bf \theta}_1)
                          + \cdots
                          + \pi_J f_J(x ; {\bf \theta}_J),
\end{eqnarray}
where $f_j(\cdot ; {\bf \theta}_j)$; $j=1, \ldots, M$, represents the pdf of $j$-th
component of  the FMM which is specified up to a vector ${\bf \theta}_j$ of unknown
parameters, known a priori  to be distinct.
The vector of all unknown parameters of the FMM \eqref{mixture} is shown by $
{\boldsymbol \Psi}= (\bpi,\bxi)$ where $\bpi=(\pi_1, \ldots, \pi_{J})$ and  ${\bxi} =(\bt_1^\top, \ldots, \bt_J)^\top$ where the superscript $\top$
stands for the vector transpose.

In the following, we describe how a ranked set sample (RSS) of size $nH$ can be constructed from an osteoporosis research example 
where $n$ and $H$ denote the set and cycle sizes, respectively. 
While measuring the response variable $X$ (i.e., BMD score) are expensive, practitioners have access to easy-to-measure 
characteristics about patients such as age, BMI or BMD scores from previous years. First, $H^2$ patients  are identified at 
random from the underlying population (without measuring their $x$-values) 
and are allocated to $H$ sets of equal size. The patients 
in the sets are ranked based on an easy-to-measure characteristic, such as age. We then select only the patients with 
$r$-th smallest rank in set $r$  to undergo the bone examination and measure their BMD scores, 
denoted by $X_{[r]1}$ for $r=1,\ldots,H$. 
The entire process is called a cycle. The cycle is then repeated $n$ times independently to collect RSS data of size 
$nH$ from the BMD population, denoted by $\{X_{[r]i}; r=1,\ldots,H; i=1,\ldots,n\}$.
In this notation, the square bracket signifies the imperfection in the ranking process.
The above RSS sampling design is balanced as we obtain the same number of observations from each rank stratum  $r=1,\ldots,H$.   

In practice, the sampling units are ranked based on values of an easy-to-measure concomitant variable (henceforth called 
ranker), which is why the declared ranks will be judgmental and may differ from the true ranks, leading to an imperfect
 RSS design. In a similar vein to \cite{hatefi2015mixture} and \cite{arslan2013parametric}, we incorporate  this imperfect ranking, 
 involving the RSS data collection, into the estimation as a missing data mechanism handled  by the misplacement probability model as 
\begin{eqnarray*}
\balpha =\left[\begin{array}{ccccc} 
                \alpha_{1,1} & \alpha_{1,2} & \cdots & \alpha_{1,H} \\
                \vdots       & \vdots & \cdots & \vdots \\
                \alpha_{H,1} & \alpha_{H,2} & \cdots & \alpha_{H,H} \\
\end{array} \right],
\end{eqnarray*}  
where $\alpha_{r,h}$  denotes the probability that units with true rank $h$ is assigned by the ranker to judgmental rank $r$.
Since $\balpha$ partition the set  into $H$ ranking strata, then $\balpha$ matrix is assumed to be a doubly stochastic matrix such that 
$\sum_{h=1}^{H} \alpha_{r,h} = \sum_{r=1}^{H} \alpha_{r,h} = 1$. 
 To incorporate the imperfect ranks of RSS data into the likelihood function, we introduce latent vector 
 $\bd_i^{[r]} = (\bd_i^{[r,1]},\ldots,\bd_i^{[r,H]})$ for each statistic $X_{[r]i}, \forall i,r$ using the misplacement probability model $\balpha$.  

The latent vector $\bd_i^{[r]}$ is designed to learn the misplacement probability model of the ranker and how  
information is traveled between the ranking strata.  
The $\bd_i^{[r]}$ are  independent  and identically distributed from a multinomial distribution with one draw out of $H$ strata with probabilities
 $\balpha^{[r]}=(\alpha_{r,1}, \ldots, \alpha_{r,H})$. Thus, the joint distribution of $(X_{[r]i},\bd_i^{[r]})$ is given by
 \begin{align}\label{f_xd}
f(x_{[r]i},\dd_i^{[r]};\bomega) = \prod_{h=1}^{H}  \left\{ \alpha_{r,h} f^{(h:H)}(x_{[r]i};\bpsi) \right\}^{\dd_i^{[r,h]}}
\end{align}
where $\bomega=(\bpsi,\balpha)$ and 
$f^{(h:H)}(\cdot;\bpsi)$ refers to the pdf of the $r$-th order statistic of \eqref{mixture} from a set of $H$ units as
\[
f^{(h:H)}(x_{[r]i};\bpsi) = H  {{H-1}\choose{h-1}} f(x_{[r]i};\bpsi) [F(x_{[r]i};\bpsi)]^{h-1} [{\bar F}(x_{[r]i};\bpsi)]^{H-h},
\]
with $F(\cdot;\bpsi)$ is the cdf of \eqref{mixture} and ${\bar F}(\cdot;\bpsi) = 1- F(\cdot;\bpsi)$.
Using \eqref{f_xd} and summing $f(x_{[r]i},\dd_i^{[r]}; \bomega) $ over $\dd_i^{[r]}$, we get the marginal distribution of $X_{[r]i}$ by
\begin{align}\label{f_xr} 
f(x_{[r]i};\bomega) =
       \sum_{\dd_i^{[r,1]}+\ldots+\dd_i^{[r,H]}=1}
            f(x_{[r]i},\dd_i^{[r]};{\bomega}) 
      =  \sum_{h=1}^{H}  \alpha_{r,h}  f^{(h:H)}(x_{[r]i};\bpsi). \end{align}
The joint distribution \eqref{f_xd}, is still not tractable in estimating the component parameters of the FMM. 
We use the missing data mechanism of \cite{hatefi2014estimation,hatefi2015mixture} and introduce three new latent
 vectors ${\bf Z}^{[r]}_i, {\bf L}^{[r]}_i$ and ${\bf U}^{[r]}_i$ for each $X_{[r]i}$ given $\bd_i^{[r]}$ to 
 incorporate the component memberships of the imperfect RSS data as an unsupervised learning approach into the likelihood function. 

Given $\bd_i^{[r]}=\dd_i^{[r]}$, the true rank of $X_{[r]i}$ is known. Let $\bd_i^{[r,h]}=1$ denote one appearing in the 
$h$-th entry of vector $\bd_i^{[r]}$. 
Let ${\bf Z}_i^{[r]} | \{\bd_i^{[r,h]}=1\}$ denote the latent vector specifying the component of $X_{[r]i}$ 
with ${\bf Z}_i^{[r]} = (Z_{i1}^{[r]},\ldots, Z_{iJ}^{[r]})$ such that
\[  Z_{ij}^{[r]}\big|\{\ld_i^{[r,h]}=1\} =
        \left\{
        \begin{array}{ll}
         1 & \mbox{if $x_{[r]i}$ belongs to component $j$}; \\
         0 & \mbox{otherwise},\end{array}
         \right.
\]
with $\sum_{j=1}^J \left( Z_{ij}^{[r]}\big|\{\ld_i^{[r,h]}=1\}\right)=1$.
Hence,  ${\bf Z}_{i}^{[r]} | \{\bd_i^{[r,h]}=1\} \sim \text{Multi}(1,\pi_1,\ldots,\pi_J)$. 
Also, given $\{\bd_i^{[r,h]}=1\}$, we introduce ${\bf L}_i^{[r]} = (L_{i1}^{[r]},\ldots, L_{iJ}^{[r]})$ 
where $L_{ij}^{[r]}|\{\bd_i^{[r,h]}=1\}$ denotes the number observations smaller than $x_{[r]i}$ coming from the 
$j$-th component of the population such that $\sum_{j=1}^J \left( L_{ij}^{[r]}\big|\{\ld_i^{[r,h]}=1\}\right)=h-1$. 
Hence,  ${\bf L}_{i}^{[r]} | \{\bd_i^{[r,h]}=1\} \sim \text{Multi}(h-1,\pi_1,\ldots,\pi_J)$. 
Finally,  given $\{\bd_i^{[r,h]}=1\}$, we introduce ${\bf U}_i^{[r]} = (U_{i1}^{[r]},\ldots, U_{iJ}^{[r]})$ 
where $U_{ij}^{[r]}|\{\bd_i^{[r,h]}=1\}$ denotes the number observations bigger than $x_{[r]i}$ coming from 
the $j$-th component with $\sum_{j=1}^J \left( U_{ij}^{[r]}\big|\{\ld_i^{[r,h]}=1\}\right)=H-h$. 
Thus,  ${\bf U}_{i}^{[r]} | \{\bd_i^{[r,h]}=1\} \sim \text{Multi}(H-h,\pi_1,\ldots,\pi_J)$. 
Owing to the fact that RSS data from FMM \eqref{mixture} are independent, the  latent variables ${\bf Z}^{[r]}_i, {\bf L}^{[r]}_i$ 
and ${\bf U}^{[r]}_i$ are conditionally independent given $\bd_i^{[r]}$. Using \eqref{f_xd}, the joint distribution of ($X_{[r]i}, \bd_i^{[r]},{\bf Z}^{[r]}_i, {\bf L}^{[r]}_i,{\bf U}^{[r]}_i$) is given by
 \begin{align}\label{f_yi} \nonumber
 &f(x_{[r]i},\dd_i^{[r]},{\bf z}_{i}^{[r]},{\bf l}_{i}^{[r]},{\bf u}_{i}^{[r]};\bomega)\\
\propto&  \prod_{h=1}^{H}   \prod_{j=1}^{J} 
 \bigg \{
    \alpha_{r,h}
    \pi_j^{\{z_{ij}^{[r]}+l_{ij}^{[r]}+u_{ij}^{[r]}\}} 
  [f_j(x_{[r]i},\theta_j)]^{z_{ij}^{[r]}}
    [F_j(x_{[r]i},\theta_j)]^{l_{ij}^{[r]}} [\bar{F}_j(x_{[r]i},\theta_j)]^{u_{ij}^{[r]}}
   \bigg \}
    ^{\ldd_i^{[r,h]}}.
 \end{align}
Note that the marginal distribution \eqref{f_xr} and the joint distribution \eqref{f_xd} can be easily retrieved  from \eqref{f_yi}
 through an marginalization step by summing over 
${\bf Z}^{[r]}_i|\bd_i^{[r]}, {\bf L}^{[r]}_i|\bd_i^{[r]}$, and ${\bf U}^{[r]}_i|\bd_i^{[r]}$.

Let ${\bf Y}_{rss}= \{(X_{[r]i}, \bd_i^{[r]},{\bf Z}^{[r]}_i, {\bf L}^{[r]}_i,{\bf U}^{[r]}_i); i=1,\ldots,n;r=1,\ldots,H\}$ denote 
the collection of the RSS data and their latent variables, henceforth called the complete RSS data.
From \eqref{f_yi}, the complete likelihood function under RSS data is given by 
\begin{align}\label{Lrss}
{\mathcal L}({\bf y}_{rss};\bomega) = \prod_{i=1}^{n} \prod_{r=1}^{H} \prod_{h=1}^{H} \prod_{j=1}^{J}
 f(x_{[r]i},\dd_i^{[r]},{\bf z}_{i}^{[r]},{\bf l}_{i}^{[r]},{\bf u}_{i}^{[r]};\bomega).
\end{align}

Let $(X_1,\ldots,X_{nH})$ denote a commonly used simple random sample (SRS) of size $nH$ from FMM \eqref{mixture}. 
It is common to introduce latent variables ${\bf Z}_i=(Z_{i1},\ldots,Z_{iJ})$ for each $X_i$ to denote the unknown component where $X_i$ comes from. 
As $\sum_{j=1}^J  Z_{ij}=1$, we have  ${\bf Z}_{i} \overset{iid}{\sim} \text{Multi}(1,\pi_1,\ldots,\pi_J)$. 
Thus, the complete likelihood function under SRS data is given by
\begin{align}\label{Lsrs}
{\mathcal L}({\bf y}_{srs};\bpsi) = \prod_{i=1}^{nH}  \prod_{j=1}^{J}
\left\{
\pi_j f_j(x_i,\theta_j) 
  \right\}^{z_{ij}}.
\end{align}

 \section{Posterior distributions from RSS data}\label{post}

According to the vital role of the finite mixture of normal distributions in mixture modelling and 
model-based classifications \citep{peel2000finite,mclachlan1988mixture} and the fact that  the goal of this manuscript 
is to compare the Bayesian proposal under RSS data with their counterparts under commonly used SRS 
data, without loss of generality, henceforth we assume that the underlying population \eqref{mixture}
 is given by a finite mixture normal distributions by
 \begin{eqnarray}\label{Nmix}
      f(x; {\bf \Psi})= \sum_{j=1}^{J} \pi_j \Phi_j(x ; \mu_j,\sigma^2_j),
 \end{eqnarray}
where $\Phi_j(\cdot ; \mu_j,\sigma^2_j)$ and $\phi_j(\cdot ; \mu_j,\sigma^2_j)$ refers, respectively, to the pdf and the cdf of normal distribution with mean $\mu_j$ and variance $\sigma^2_j$.

For the sake of completeness, we first briefly describe the Gibbs sampling under SRS data from mixture population \eqref{Nmix}. Following \cite{casella2021statistical} and \cite{robert2007bayesian},
one can use conjugate priors for the parameters by
\begin{align}\label{het_prior}
\sigma_j^2 \sim \text{IG}(\nu_j,\beta_j), ~ \mu_j| \sigma_j^2 \sim N(\kappa_j, \sigma_j^2 / \tau_j),  ~  {\bf\pi} \sim {\mathcal D}(\gamma_1,\ldots,\gamma_J),
\end{align}
where $\text{IG}$ and ${\mathcal D}$ denote the inverse Gamma and Dirichlet distribution, respectively. Note that  $(\nu_j,\beta_j,\kappa_j,\tau_j,\gamma_j)$ denotes  the set of fixed hyper-parameters of the model for $j=1,\ldots,J$. 
Let $\bpsi^{(0)}$ be the starting point. Also let  $\bpsi^{(t)}$ and ${\bf Z}^{(t)}$  be the updates from the $t$-th iteration.   Using \eqref{Lsrs} and \eqref{het_prior}, we compute  the $(t+1)$-th iteration of the SRS-based Gibbs sampler as follows:
\begin{align} \nonumber \label{mu_het_srs}
\P(Z_{ij}^{(t+1)}| \{\bpsi^{(t)}, {\bf y}^{(t)}_{srs}\}) &\propto \frac{\pi_j^{(t)}}{\sigma_j^{(t)}} \, 
\exp{\left( -\frac{(x_i-\mu_j^{(t)})^2}{2 (\sigma_j^2)^{(t)}} \right)}, \\\nonumber
{\bf \pi}^{(t+1)} | \{\bpsi^{(t)}, {\bf y}^{(t+1)}_{srs}\} &\sim {\mathcal D}(n_1^{(t+1)}+\gamma_1,\ldots,n_j^{(t+1)}+\gamma_J),\\
\mu_j^{(t+1)}| \{\bpsi^{(t)}, {\bf y}^{(t+1)}_{srs}\} & \sim N\left(\frac{\tau_j \kappa_j + S_{1,j}^{(t+1)}}{\tau_j + n_j^{(t+1)}}, \frac{(\sigma_j^2)^{(t)}}{\tau_j + n_j^{(t+1)}}\right),
\end{align}
where $n_j^{(t+1)} = \sum_{i=1}^{nH} \I(Z_{ij}^{(t+1)}=1)$ and 
$S_{1,j}^{(t+1)}= \sum_{i=1}^{nH} x_i \I(Z_{ij}^{(t+1)}=1)$. We then compute $S_{2,j}^{(t+1)}= \sum_{i=1}^{nH} (x_i-\mu_j^{(t+1)})^2 \I(Z_{ij}^{(t+1)}=1) $ and  update 
 \begin{align} \label{sig_het_srs}
(\sigma_j^2)^{(t+1)}| \{\bpsi^{(t+1)}, {\bf y}^{(t+1)}_{srs}\} \sim \text{IG}\left( \nu_j+\frac{1}{2}(n_j^{(t+1)}+1),\beta_j+\frac{\tau_j}{2}(\mu_j^{(t+1)}-\kappa_j)^2
+ \frac{S_{2,j}^{(t+1)}}{2}\right).
\end{align}

The mixture population \eqref{Nmix} is considered homosedastic when  the component densities  have an 
equal variance; that is $\sigma_1^2=\ldots=\sigma_J^2=\sigma^2$. In this case, one can use \cite{casella2021statistical} and  \cite{robert2007bayesian} and introduce homosedastic version of the priors of \eqref{het_prior} by 
\begin{align}\label{hom_prior}
\sigma^2 \sim \text{IG}(\nu,\beta), ~ \mu_j| \sigma^2 \sim N(\kappa_j, \sigma^2 / \tau_j),  ~  {\bf\pi} \sim {\mathcal D}(\gamma_1,\ldots,\gamma_J),
\end{align}
where  $(\nu,\beta,\kappa_j,\tau_j,\gamma_j)$ denotes the set of fixed hyper-parameters of the model for $j=1,\ldots,J$. 
Like the heteroscedastic case, the $(t+1)$-th iteration of the SRS-based Gibbs sampler under mixture of homosedastic normals is given by
\begin{align} \nonumber \label{mu_hom_srs}
\P(Z_{ij}^{(t+1)}| \{\bpsi^{(t)}, {\bf y}^{(t)}_{srs}\}) &\propto \frac{\pi_j^{(t)}}{\sigma^{(t)}} \, 
\exp{\left( -\frac{(x_i-\mu_j^{(t)})^2}{2 (\sigma^2)^{(t)}} \right)}, \\\nonumber
{\bf \pi}^{(t+1)} | \{\bpsi^{(t)}, {\bf y}^{(t+1)}_{srs}\} &\sim {\mathcal D}(n_1^{(t+1)}+\gamma_1,\ldots,n_j^{(t+1)}+\gamma_J),\\
\mu_j^{(t+1)}| \{\bpsi^{(t)}, {\bf y}^{(t+1)}_{srs}\} & \sim N\left(\frac{\tau_j \kappa_j + S_{1,j}^{(t+1)}}{\tau_j + n_j^{(t+1)}}, \frac{(\sigma^2)^{(t)}}{\tau_j + n_j^{(t+1)}}\right), \\\label{sig_hom_srs}
(\sigma^2)^{(t+1)}| \{\bpsi^{(t+1)}, {\bf y}^{(t+1)}_{srs}\} &\sim \text{IG}\left( \nu+\frac{1}{2}(nH+J),\beta+\frac{\tau_j}{2}(\mu_j^{(t+1)}-\kappa_j)^2
+ \frac{S_{2,j}^{(t+1)}}{2}\right),
\end{align}
where $n_j^{(t+1)},S_{1,j}^{(t+1)}$ and $S_{2,j}^{(t+1)}$ are computed similar to the heteroscedastic SRS-based Gibbs sampler. 

In the rest of this section, we develop the posterior distributions for the mixture parameters using the imperfect RSS data. 
According to the unique structure of the complete RSS data, we have to introduce  a new Gibbs sampler to augment the latent
 variable consisting of ranking information and component memberships.   
The misplacement probabilities $\balpha$ control the imperfect ranking information travelling between rank strata. There is no such concept and parameters in SRS data. Therefore, throughout this manuscript, we assume that the RSS design's ranking parameter $\balpha$ is fixed and should be estimated in the algorithm based on collected  RSS data. To do so, we develop a separate EM algorithm \citep{dempster1977maximum} to encompass the estimation and maximization steps of these missing values within the RSS Bayesian estimation procedure. 
 
 We first need to find the marginal distributions of latent variables given RSS data.
 From \eqref{f_yi} and \eqref{f_xr}, it is easy to show that 
 $\bd_i^{[r]}|X_{[r]i} \sim \text{Multi}(1,\zeta_i^{[r,1]},\ldots,\zeta_i^{[r,H]})$ where
 \begin{eqnarray}\label{zeta}
\zeta_{i}^{[r,h]}(\bomega)=
\frac
{\alpha_{r,h} { B}_{h,H-h+1}(F(x_{[r]i};{\bf\Psi}))}
{\displaystyle \sum_{h'=1}^H	  \alpha_{r,h'}
{ B}_{h',H-h'+1}(F(x_{[r]i};{\bf\Psi}))},
\end{eqnarray}
 where $F(\cdot;{\bf\Psi})$ is the cdf of \eqref{Nmix} and $B_{a,b}(\cdot)$  denotes the pdf of Beta distribution with parameters $a$ and $b$. 
 To augment the component membership of RSS-based latent variables in the Gibbs sampling, one can easily obtain the conditional distributions 
 ${\bf Z}_i^{[r]},{\bf L}_i^{[r]}$ and ${\bf U}_i^{[r]}$ given $X_{[r]i},\bd_i^{[r,h]}=1$ by
 \begin{eqnarray}\label{fz|xd}
{\bf Z}_{i}^{[r]}|\{x_{[r]i},\ld_i^{[r,h]}=1\}\sim \text{Multi}
\left(
1,\frac{ \pi_1 \Phi_1(x_{[r]i} ; \mu_1,\sigma^2_1)}{f(x_{[r]i};{\bf{\Psi}})},\ldots, \frac{ \pi_J \Phi_J(x_{[r]i} ; \mu_J,\sigma^2_J)}{f(x_{[r]i};{\bf{\Psi}})}
\right),
\end{eqnarray}
\begin{eqnarray}\label{fl|xd}
{\bf L}_{i}^{[r]}|\{x_{[r]i},\ld_i^{[r,h]}=1\}\sim \text{Multi}
\left(
h-1,\frac{ \pi_1 \phi_1(x_{[r]i} ; \mu_1,\sigma^2_1)}{F(x_{[r]i};{\bf{\Psi}})},\ldots, \frac{ \pi_J \phi_J(x_{[r]i} ; \mu_J,\sigma^2_J)}{F(x_{[r]i};{\bf{\Psi}})}
\right),
\end{eqnarray}
\begin{eqnarray}\label{fu|xd}
{\bf U}_{i}^{[r]}|\{x_{[r]i},\ld_i^{[r,h]}=1\}\sim \text{Multi}
\left(
H-h,\frac{ \pi_1 {\bar\phi}_1(x_{[r]i} ; \mu_1,\sigma^2_1)}{{\bar F}(x_{[r]i};{\bf{\Psi}})},\ldots, \frac{ \pi_J {\bar\phi}_J(x_{[r]i} ; \mu_J,\sigma^2_J)}{{\bar F}(x_{[r]i};{\bf{\Psi}})}
\right),
\end{eqnarray}
where ${\bar\phi} = 1- {\phi}$ and ${\bar F} = 1-F$.
 
  Owing to the doubly stochastic property of the misplacement probability matrix, one can translate  
  maximizing the logarithm of \eqref{Lrss} to a constraint optimization problem using the Lagrangian 
  multiplayers ${\bf\lambda}=(\lambda_1,\ldots,\lambda_H)$ and ${\bf\lambda}'=(\lambda_1',\ldots,\lambda_H')$. 
  The maximization step is then given by
 \begin{align}\label{ma_step}
 Q(\balpha,\bomega;{\bf\lambda},{\bf\lambda}') 
 = \sum_{h=1}^{H} \sum_{h'=1}^{H} \zeta_{h,h'}(\bomega) \log(\alpha_{h,h'})
 + \sum_{h=1}^{H} \lambda_h \left\{ \sum_{h'=1}^{H} \alpha_{h,h'} -1 \right\}  
  + \sum_{h=1}^{H} \lambda_{h'} \left\{ \sum_{h'=1}^{H} \alpha_{h',h} -1 \right\},  
 \end{align}
where   $\zeta_{r,h}(\bomega) = \sum_{i=1}^{n}  \zeta_i^{[r,h]}(\bomega)$ from \eqref{zeta}. For more details  about the 
above maximization algorithm, readers are referred to \cite{arslan2013parametric}.

Since RSS data, in the absence of ranking information, can be considered as managed SRS data from the same population, we follow \cite{casella2021statistical} and \cite{robert2007bayesian}  and propose the prior distributions  \eqref{het_prior} under RSS data. From complete 
 likelihood function \eqref{Lrss} and \eqref{het_prior}, we develop a Metropolis-within-Gibbs sampling approach to find the posterior distributions of the mixture parameters under imperfect RSS design. Let $\bomega^{(0)} =(\bpsi^{(0)},\balpha^{(0)})$ be the stating point. Also let $(\bomega^{(t)}, {\bf Z}^{(t)}, {\bf L}^{(t)},{\bf U}^{(t)})$ be the update from the $t$-th iteration of the Gibbs sampler. Accordingly, the $(t+1)$-th iteration of the  RSS-based Metropolis-within-Gibbs sampler is developed as follows.
 
\noindent{\bf EM Step}: As the ranking parameters are treated fixed; hence, we first apply \eqref{zeta} and obtain
$(\zeta_{i}^{[r,h]})^{(t)} = \zeta_{i}^{[r,h]}(\bomega^{(t)})$ in E-step of the algorithm. In the M-step,
we update $\balpha^{(t+1)}$ from the constraint optimization \eqref{ma_step} by $\balpha^{(t+1)} = \arg \max_{\balpha}  Q(\balpha,\bomega^{(t)};{\bf\lambda},{\bf\lambda}') $ using $(\zeta_{i}^{[r,h]})^{(t)}$ calculated in the E-step.

\noindent{\bf Augmentation Step}: In this step, we use  \eqref{zeta} - \eqref{fu|xd} to  sample from the conditional distribution of latent variables for $r=1,\ldots,H;i=1,\ldots,n$.
 We first employ \eqref{zeta} to sample  $(\bd_i^{[r]})^{(t+1)}$ from  distribution 
 $(\bd_i^{[r]})| \{\balpha^{(t+1)},\bpsi^{(t)},{\bf y}_{rss}^{(t)}\} \sim 
 \text{Multi}\left(1,(\zeta_i^{[r,1]})^{(t+1)},\ldots,(\zeta_i^{[r,1]})^{(t+1)}\right)$ where $(\zeta_i^{[r,h]})^{(t+1)}=\zeta_i^{[r,h]}(\bomega)|_{\bomega=(\balpha^{(t+1)},\bpsi^{(t)})}$. 
 From \eqref{fz|xd}, we sample 
$({\bf Z}_i^{[r]})^{(t+1)}| \{x_{[r]i},(\bd_i^{[r]})^{(t+1)},\bpsi^{(t)}\} \sim 
 \text{Multi}\left(1,p^z_1,\ldots,p^z_J\right)$ where 
\[
p^z_j =  \pi_j^{(t)} \Phi_j(x_{[r]i} ; \mu_j^{(t)},(\sigma^2_j)^{(t)}) / f(x_{[r]i};{\bf{\Psi}}^{(t)}), ~ j=1,\ldots,J.
\]
We use  \eqref{fl|xd} and generate a sample from
$({\bf L}_i^{[r]})^{(t+1)}| \{x_{[r]i},(\bd_i^{[r]})^{(t+1)},\bpsi^{(t)}\} \sim 
 \text{Multi}\left(1,p^l_1,\ldots,p^l_J\right)$ with 
\[
p^l_j =  \pi_j^{(t)} \phi_j(x_{[r]i} ; \mu_j^{(t)},(\sigma^2_j)^{(t)}) / F(x_{[r]i};{\bf{\Psi}}^{(t)})~ j=1,\ldots,J.
\]
We apply \eqref{fu|xd}  to sample $({\bf U}_i^{[r]})^{(t+1)}$ from 
$({\bf U}_i^{[r]})| \{x_{[r]i},(\bd_i^{[r]})^{(t+1)},\bpsi^{(t)}\} \sim 
 \text{Multi}\left(1,p^u_1,\ldots,p^u_J\right)$ with 
\[
p^u_j =  \pi_j^{(t)} {\bar\phi}_j(x_{[r]i} ; \mu_j^{(t)},(\sigma^2_j)^{(t)}) / {\bar F}(x_{[r]i};{\bf{\Psi}}^{(t)})~ j=1,\ldots,J.
\] 

\noindent{\bf $\pi$-Step}: Let ${\bf Y}_{rss}^{(t+1)}= \left\{(X_{[r]i}, (\bd_i^{[r]})^{(t+1)},({\bf Z}^{[r]}_i)^{(t+1)}, ({\bf L}^{[r]}_i)^{(t+1)},
({\bf U}^{[r]}_i)^{(t+1)}); \forall i,r\right\}$ from the data augmentation step.  We update the posterior distribution of the mixing proportions  by
$
{\bf \Pi}^{(t+1)} | \{\bpsi^{(t+1)}, {\bf y}^{(t+1)}_{rss}\} \sim {\mathcal D}(\gamma_1^*,\ldots,\gamma_J^*)
$
where 
\[
\gamma_j^*= \sum_{i=1}^{n} \sum_{r=1}^{H} \sum_{h=1}^{H} (\bd_i^{[r,h]})^{(t+1)}
\left\{ ({\bf Z}_i^{[r]})^{(t+1)} + ({\bf L}_i^{[r]})^{(t+1)} + ({\bf U}_i^{[r]})^{(t+1)} \right\} + \gamma_j.
\]

\noindent{\bf $\xi$-Step}: One can easily show that the univariate posterior distribution of the component parameters of \eqref{Nmix} are given 
 \begin{align} \label{mu_het_g_rss} \nonumber
g_{rss}\left(\mu_j|\{ {\bf y}^{(t+1)}_{rss},(\sigma_j^2)^{(t)}\}\right) &\propto
      \prod_{i=1}^{n} \prod_{r=1}^{H} \prod_{h=1}^{H} 
       \left\{
       \left[
       \Phi_j(x_{[r]i} ; \mu_j,(\sigma^2_j)^{(t)})
       \right]^{({z_{ij}^{[r]}})^{(t+1)}}
       \right. \\\nonumber
   &  \left.
       \times
       \left[
        \phi_j(x_{[r]i} ; \mu_j,(\sigma^2_j)^{(t)})
       \right]^{({l_{ij}^{[r]}})^{(t+1)}}
       \left[
        {\bar\phi}_j(x_{[r]i} ; \mu_j,(\sigma^2_j)^{(t)})
       \right]^{({u_{ij}^{[r]}})^{(t+1)}}
        \right
        \}^{(\ldd_i^{[r,h]})^{(t+1)}}\\
   &  \times \frac{\sqrt{\tau_j}}{\sigma_j^{(t)}} \exp\left(- \frac{\tau_j (\mu_j-\kappa_j)^2}{2  (\sigma_j^2)^{(t)}}\right),    
 \end{align}
 
  \begin{align} \label{sigma_het_g_rss} \nonumber
g_{rss}\left(\sigma^2_j|\{ {\bf y}^{(t+1)}_{rss},\mu_j^{(t+1)}\}\right) &\propto
      \prod_{i=1}^{n} \prod_{r=1}^{H} \prod_{h=1}^{H} 
       \left\{
       \left[
       \Phi_j(x_{[r]i} ; \mu_j^{(t+1)},\sigma^2_j)
       \right]^{({z_{ij}^{[r]}})^{(t+1)}}
       \left[
        \phi_j(x_{[r]i} ; \mu_j^{(t+1)},\sigma^2_j)
       \right]^{({l_{ij}^{[r]}})^{(t+1)}} 
       \right. \\\nonumber
    &  \left.
       \times
       \left[
        {\bar\phi}_j(x_{[r]i} ; \mu_j^{(t+1)},\sigma^2_j)
       \right]^{({u_{ij}^{[r]}})^{(t+1)}}
        \right
        \}^{(\ldd_i^{[r,h]})^{(t+1)}}
   \frac{\sqrt{\tau_j}}{\sigma_j} \exp\left(-\frac{\tau_j (\mu_j^{(t+1)}-\kappa_j)^2}{2  \sigma^2_j}\right)\\
    &  \times   \frac{\beta_j^{\nu_j}}{\Gamma(\nu_j)} \left( \frac{1}{\sigma^2_j} \right)^{\nu_j+1} \exp\left(-\frac{\beta_j}{\sigma^2_j} \right).
 \end{align}
As it is obvious from \eqref{mu_het_g_rss} and \eqref{sigma_het_g_rss}, unlike the SRS-based Gibbs sampling, 
there is no closed form for the posterior distribution of the component parameters.
To cope with this challenge, we propose a Metropolis-Hastings algorithm within the RSS Gibbs sampler \citep{robert1999monte,robert2007bayesian} to take samples 
from \eqref{mu_het_g_rss} and \eqref{sigma_het_g_rss}. 

Unlike the Gibbs sampling approach, the Metropolis-Hastings algorithm produces dependent  samples from the target 
posterior density by generating candidates from an instrumental density.  The generated candidate is then accepted 
as the next state of the MCMC chain based on an acceptance probability. 
To generate candidates from the target posterior distribution \eqref{mu_het_g_rss}, we treated the posterior density 
\eqref{mu_het_srs} using RSS data, say $g_{srs}\left(\mu_j|\{ {\bf y}^{(t+1)}_{rss},(\sigma_j^2)^{(t)}\}\right)$, as 
the instrumental density and generate the candidate $\mu_j^*$. 
Through an stochastic step, the candidate  is accepted as $\mu_j^{(t+1)}$ with probability
\begin{align} \label{acc_prob}
\rho(\mu_j^*,\mu_j^{(t+1)}) = \min\left( \frac{g_{rss}\left(\mu_j^*|\{ {\bf y}^{(t+1)}_{rss},(\sigma_j^2)^{(t)}\}\right) g_{srs}\left(\mu_j^{(t)}|\{ {\bf y}^{(t+1)}_{rss},(\sigma_j^2)^{(t)}\}\right)}{g_{rss}\left(\mu_j^{(t)}|\{ {\bf y}^{(t+1)}_{rss},(\sigma_j^2)^{(t)}\}\right) g_{srs}\left(\mu_j^*|\{ {\bf y}^{(t+1)}_{rss},(\sigma_j^2)^{(t)}\}\right)},1 \right).
\end{align}
 Similarly, we apply the  Metropolis-Hastings algorithm to generate $(\sigma^2_j)^{(t+1)}$ from \eqref{sigma_het_g_rss} 
 using the SRS-based posterior distribution \eqref{sig_het_srs} under RSS data as the instrumental density.
 
 One can generalize the above RSS-based Gibbs sampler to the case where the mixture population \eqref{Nmix} comprises  
 homosedastic normal distributions, that is $\sigma_1=\ldots=\sigma_J=\sigma$.
 Using the prior distributions \eqref{hom_prior} 
  and the complete RSS likelihood function \eqref{Lrss} (under homosedastic mixture model), we develop a Metropolis-within-Gibbs sampler for the imperfect RSS data from mixture of homosedstic normal distributions. 
Let $\bomega^{(0)}$ and  $(\bomega^{(t)}, {\bf y}_{rss}^{(t)})$ denote the starting point and the updates from the $t$-th iteration of Gibbs sampler, respectively.
Then the $(t+1)$-th iteration of the  RSS-based Gibbs sampler in estimating the parameters of the homosedastic population is developed as follows: The EM, Augmentation  and ${\bf\pi}$- Steps remain the same as described earlier, applying the homosedastic property of  the underlying mixture population. 

 \noindent{\bf $\xi$-Step}: From  the priors \eqref{hom_prior} and \eqref{Lrss}, the univariate posterior distributions of $\mu_j$ and $\sigma^2$ are given by
 \begin{align} \label{mu_hom_g_rss} \nonumber
g_{rss}\left(\mu_j|\{ {\bf y}^{(t+1)}_{rss},(\sigma^2)^{(t)}\}\right) &\propto
      \prod_{i=1}^{n} \prod_{r=1}^{H} \prod_{h=1}^{H} 
       \left\{
       \left[
       \Phi_j(x_{[r]i} ; \mu_j,(\sigma^2)^{(t)})
       \right]^{({z_{ij}^{[r]}})^{(t+1)}}
       \right. \\\nonumber
   &  \left.
       \times
       \left[
        \phi_j(x_{[r]i} ; \mu_j,(\sigma^2)^{(t)})
       \right]^{({l_{ij}^{[r]}})^{(t+1)}}
       \left[
        {\bar\phi}_j(x_{[r]i} ; \mu_j,(\sigma^2)^{(t)})
       \right]^{({u_{ij}^{[r]}})^{(t+1)}}
        \right
        \}^{(\ldd_i^{[r,h]})^{(t+1)}}\\
   &  \times \frac{\sqrt{\tau_j}}{\sigma^{(t)}} \exp\left(- \frac{\tau_j (\mu_j-\kappa_j)^2}{2  (\sigma^2)^{(t)}}\right),    
 \end{align}
 
  \begin{align} \label{sigma_hom_g_rss} \nonumber
g_{rss}\left(\sigma^2|\{ {\bf y}^{(t+1)}_{rss},\mu_j^{(t+1)}\}\right) &\propto
      \prod_{i=1}^{n} \prod_{r=1}^{H} \prod_{h=1}^{H} \prod_{j=1}^{J}
       \left\{
       \left[
       \Phi_j(x_{[r]i} ; \mu_j^{(t+1)},\sigma^2)
       \right]^{({z_{ij}^{[r]}})^{(t+1)}}
       \left[
        \phi_j(x_{[r]i} ; \mu_j^{(t+1)},\sigma^2)
       \right]^{({l_{ij}^{[r]}})^{(t+1)}} 
       \right. \\\nonumber
    &  \left.
       \times
       \left[
        {\bar\phi}_j(x_{[r]i} ; \mu_j^{(t+1)},\sigma^2)
       \right]^{({u_{ij}^{[r]}})^{(t+1)}}
        \right
        \}^{(\ldd_i^{[r,h]})^{(t+1)}}
   \prod_{j=1}^{J} 
   \left\{ \frac{\sqrt{\tau_j}}{\sigma} \exp\left(- \frac{\tau_j (\mu_j^{(t+1)}-\kappa_j)^2}{2  \sigma^2}\right)
   \right\}\\
    &  \times   \frac{\beta^{\nu}}{\Gamma(\nu)} \left( \frac{1}{\sigma^2} \right)^{\nu+1} \exp\left(-\frac{\beta}{\sigma^2} \right).
 \end{align}
 Like hetrosedastic case, we exploit Metropolis-Hastings algorithm  to sample from  \eqref{mu_hom_g_rss} and \eqref{sigma_hom_g_rss}. Here we first use Metropolis-Hastings algorithm to generate candidates $(\mu_1^*,\ldots,\mu_J^*,\sigma)$ from $g_{srs}\left(\mu_j|\{ {\bf y}^{(t+1)}_{rss},(\sigma^2)^{(t)}\}\right)$ in \eqref{mu_hom_srs}  
  and  
 $g_{srs}\left(\sigma^2|\{ {\bf y}^{(t+1)}_{rss},\mu_1^{(t+1)}, \ldots, \mu_J^{(t+1)}\}\right)$ in \eqref{sig_hom_srs}, 
 treating the RSS data as SRS data by ignoring their ranking information. The proposed candidates are then accepted through a stochastic step with the acceptance probability \eqref{acc_prob} using the target posterior distribution \eqref{mu_hom_g_rss} (and posterior distribution \eqref{sigma_hom_g_rss} for $\sigma$) with $g_{srs}\left(\mu_j|\{ {\bf y}^{(t+1)}_{rss},(\sigma^2)^{(t)}\}\right)$ $\left(\text{and $g_{srs}\left(\sigma^2|\{ {\bf y}^{(t+1)}_{rss},\mu_1^{(t+1)}, \ldots, \mu_J^{(t+1)}\}\right)$ for $\sigma$}\right)$ as the instrumental density for $\mu_j; j=1,\ldots,J$. 

 \section{Simulation Studies}\label{sim}

In this section, we compare the performance of the developed RSS-based Metropolis-within-Gibbs 
sampler with the commonly 
used SRS-based Gibbs sampler in estimating the parameters
of a finite mixture of normal distributions. We present two simulation studies to investigate 
the effect of sample size, ranking ability and number of components of the finite mixture 
models on the performance of the Bayesian estimation methods. Both simulation studies 
consist of two stages. In the first stage, we simulate the ranking misplacement probabilities 
for a given ranking ability. The probabilities are then treated as the true ranking parameters 
of the imperfect RSS sampling design to obtain the ML estimates of $\balpha$ and Bayesian 
estimates of $\bpsi$ in the second 
stage of the simulation studies.

\begin{figure}[ht]
\centering
\includegraphics[width=0.75\textwidth]{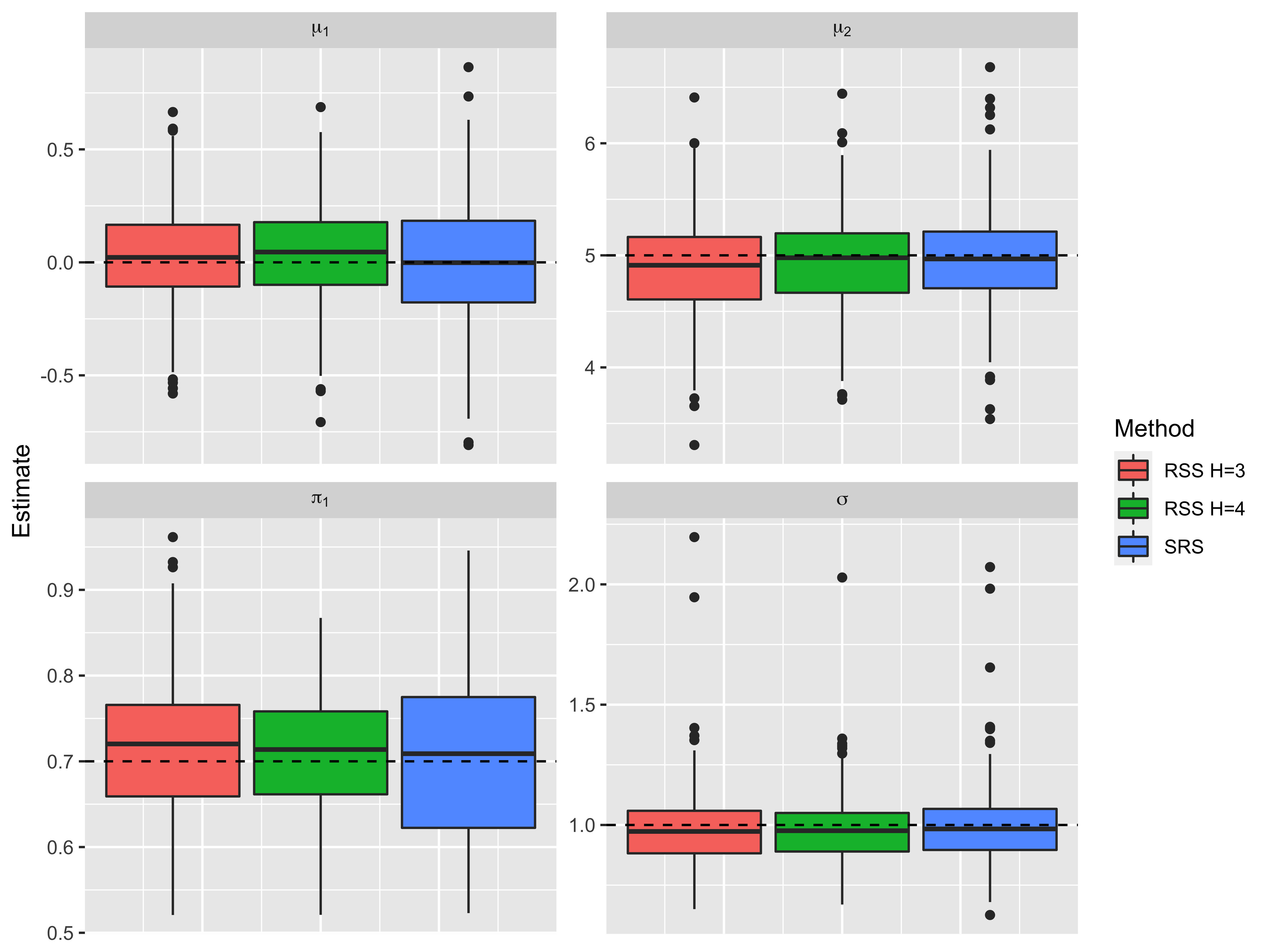}
\caption{\footnotesize{The box-plots for the Bayesian estimates of the mixture parameters under SRS and RSS samples of size $N=24$ with $H=\{3,4\}$ and $\rho=0.9$.}}
 \label{r09_05}
\end{figure}

In the first study, we simulate the RSS and SRS data of size $nH=24$ for $H=\{3,4\}$ with 
ranking ability $\rho=\{0.7,0.9\}$  from the population comprising a mixture of two homoscedastic 
normal distributions with true parameters  $\bpsi_0=(\pi_1,\mu_1,\mu_2,\sigma)=(0.7,0.,5,1)$. In the first 
stage, we have to find the misplacement probability model $\balpha$ 
corresponding to $\rho$. Let ${\bf C}$ denote a zero matrix of size $(H\times H)$. We select a 
set of $H$ observations 
from the mixture population. Following \cite{dell1972ranked}, we produce ranking concomitant 
variable $Z_i = X_i + \epsilon_i$ with $\epsilon_i \overset{iid}{\sim} N(0,\frac{1-\rho^2}{\rho^2} \sigma^2)$ 
for $i=1,\ldots,H$  such that $X$ and $Z$ have the correlation $\rho=cor(X,Z)$.  Units are ranked 
based on their $Z$-values, and the unit with judgmental rank $r$ is then selected. Let $h$ denote 
the true rank of the selected unit. We then update the $(r,h)$-the entry of ${\bf C}$, namely 
${\bf C}_{[r,h]}$ by ${\bf C}_{[r,h]} = {\bf C}_{[r,h]} +1$.
We replicate the process 5000 times and estimate $\balpha= {\bf C}/5000$.  The third column of 
Table \ref{alpha} shows the misplacement probabilities computed from the first stage. As 
$\balpha$ is a doubly stochastic matrix, Table \ref{alpha} shows only the three independent 
parameters $\balpha$ for $H=3$, as the true parameters, to highlight the fact that these 
probabilities are then treated as the true values of the ranking parameters in the second 
stage of the simulation study.  

In the second stage, we compare the proposed RSS-based Metropolis-within-Gibbs sampling 
performance with the SRS-based Gibbs sampling in estimating the population parameters. 
Throughout this paper, we used the mode of the posterior distributions to estimate the 
mixture model parameters. We used the conjugate prior distributions \eqref{hom_prior} to 
initialize the Gibbs samplers. First, we separately applied a single  K-means step to SRS 
and RSS data (ignoring the ranking information) and obtained their labelled data $y_i,i=1,\ldots,nH$. 
Following \cite{raftery1995hypothesis}, we then used the labelled data to compute the 
data-dependent hyper-parameters  of the priors  by 
$\kappa_j = {\bar y}_j, \tau_j = 2.6/(y_{j,max}-y_{j,min})^2,\nu=1.28$ and $\beta=0.36 s_y^2$.

\begin{table}[ht]
      \caption{\footnotesize{The $10$ (L), $50$ (M), and $90$ (U) percentiles of squared errors and $2.5$ (L), $50$ (M) and $97.5$ (U) percentiles for length of the 95\% shortest credible interval with coverage probabilities for the Bayesian estimates of the mixture parameters under SRS and RSS samples of size $N=24$ with $H=\{3,4\}$ and $\rho=0.9$. }}
\vspace{0.3cm} 
\centering 
{\footnotesize
\begin{tabular}{c c c c c c c c c c c c c c c}
\hline
\hline
    \multirow{ 2}{*}{Method} &  & \multirow{ 2}{*}{H} & &  \multirow{ 2}{*}{Estimand} & &
    \multicolumn{3}{c}{Squared Error} &  &
    \multicolumn{3}{c}{CI Width} & & \multirow{2}{*}{Coverage} 
    \\
    \cline{7-9}
    \cline{11-13}
     & & & & & & L & M & U & & L & M & U & &\\
    \hline
    \multirow{4}{*}{SRS} & & \multirow{4}{*}{-} & & $\pi_1$ & & 
    0.000 & 0.006 & 0.028 && 0.251 & 0.316 & 0.355 && 0.993 \\  
    & & & & $\mu_1$ & & 
    0.001 & 0.033 & 0.189 && 0.888 & 2.568 & 6.091 && 0.960 \\
    & & & & $\mu_2$ & & 
    0.003 & 0.067 & 0.498 && 1.674 & 5.167 & 6.435 && 0.964 \\  
    & & & & $\sigma$ & & 
    0.000 & 0.008 & 0.053 && 0.491 & 0.629 & 1.138 && 0.964 \\ 
    \hline
    \multirow{4}{*}{RSS} & & \multirow{4}{*}{3} & & $\pi_1$ & & 
    0.000 & 0.003 & 0.022 && 0.249 & 0.285 & 0.310 && 0.950 \\ 
    & & & & $\mu_1$ & & 
    0.000 & 0.023 & 0.137 && 0.703 & 0.918 & 5.400 && 0.943 \\  
    & & & & $\mu_2$ & & 
    0.002 & 0.086 & 0.602 && 1.348 & 1.914 & 5.919 && 0.936 \\ 
    & & & & $\sigma$ & & 
    0.000 & 0.008 & 0.047 && 0.324 & 0.399 & 0.610 && 0.867 \\  
    \hline
        \multirow{4}{*}{RSS} & & \multirow{4}{*}{4} & & $\pi_1$ & & 
    0.000 & 0.003 & 0.020 && 0.236 & 0.274 & 0.303 && 0.957 \\ 
    & & & & $\mu_1$ & & 
    0.001 & 0.021 & 0.132 && 0.680 & 0.894 & 5.345 && 0.957 \\  
    & & & & $\mu_2$ & & 
    0.002 & 0.066 & 0.392 && 1.298 & 1.800 & 5.907 && 0.955 \\ 
    & & & & $\sigma$ & & 
    0.000 & 0.007 & 0.038 && 0.325 & 0.397 & 0.548 && 0.893 \\ 
    \hline
    \hline
    \end{tabular}
    }
    \label{2_comp_rho09}
\end{table}
 
Because ranking probabilities of $\balpha$ are treated as fixed and unknown parameters, 
the RSS-based Gibbs sampler requires an EM algorithm in each iteration of the Gibbs 
sampling to estimate the parameters of the misplacement probability model. We initialized 
the EM algorithm step with a random ranking assignment  
  $\{\alpha_{r,h}^{(0)}=\frac{1}{H}, \forall r,h=1,\ldots,H\}$. We followed 
  \cite{arslan2013parametric} and \cite{hatefi2015mixture} to implement the constraint optimization 
  of the EM algorithm with stopping rule 
 $|| \balpha^{(t+1)} - \balpha^{(t)} ||_{\infty} \le 10^{-7}$ where $||\cdot||_\infty$ 
  denotes the maximum absolute value of the vector.
 The maximum number of iterations of the EM algorithm was set to 100. The entire Gibbs 
 sampling is stopped if the EM algorithm fails to converge. 
 As described in Section \ref{like}, we then implemented the Augmentation, $\Pi$- and $\xi$- 
 steps of the Gibbs sampler to update the next state of the MCMC chains. We ran SRS and RSS-based 
 Gibbs samplings for 15000 iterations. In order to wash out the effect of the initialization step 
 on the MCMC chains, we applied the burn-in period and threw away the first 5000 states. According 
 to Markovian property, the MCMC chain leads to dependent samples. Thinning is a common approach 
 to reduce the dependence between the chain states to achieve some independent samples. Thinning 
 suggests taking a sample out of every $k$ states of the MCMC chain. To do so, we applied 
a thinning step with $k=5$ to both SRS- and RSS-based Gibbs samplings.  
We estimated the mixture parameters by the mode of the posterior distributions using the SRS and 
RSS-based Gibbs samplings. The RSS ranking parameters are eventually estimated by the results of 
the EM algorithm in the last state of the MCMC chain. 
 
 \begin{figure}
\centering
\includegraphics[width=1\textwidth]{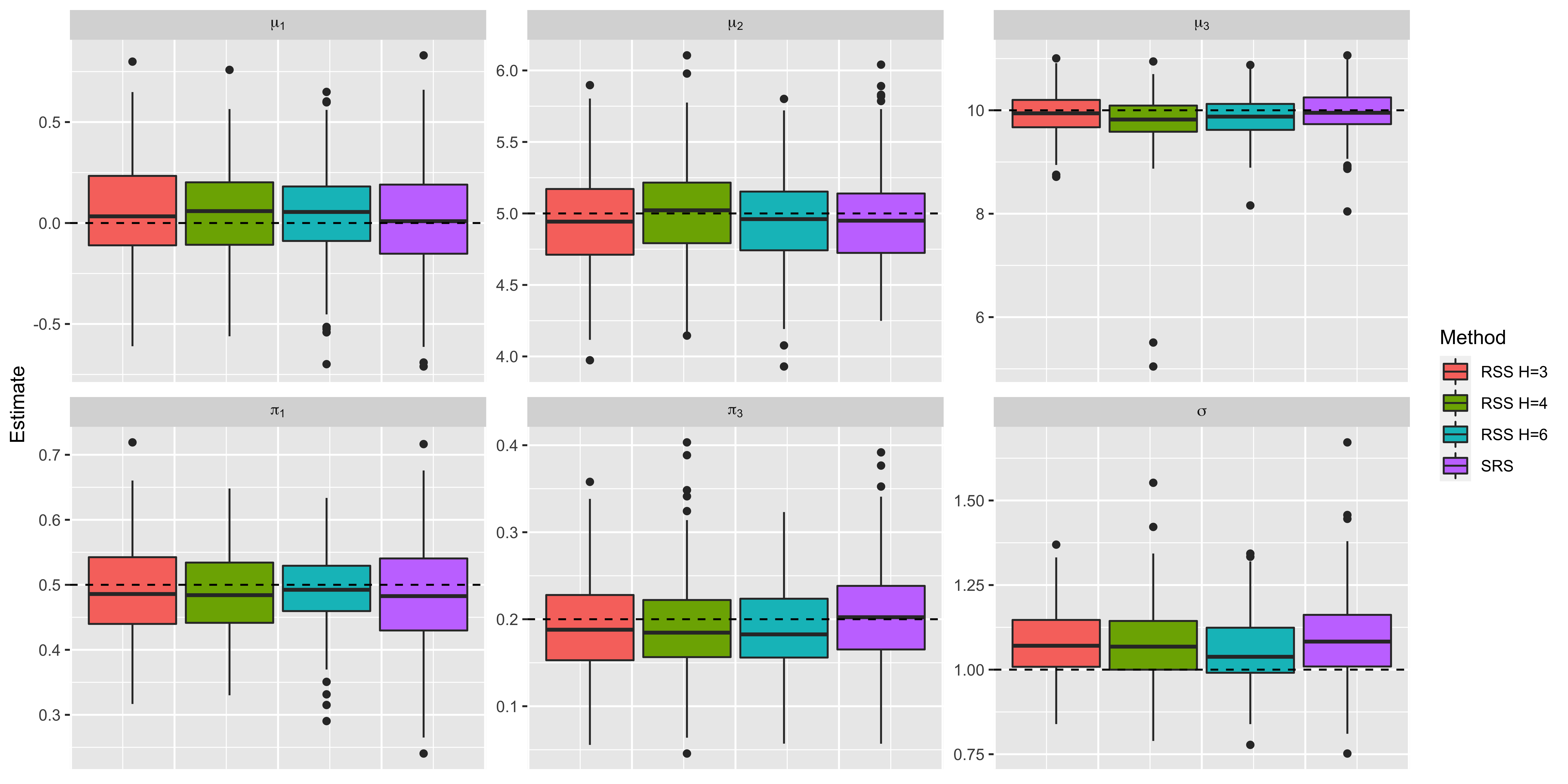}
\caption{\footnotesize{The box-plots for the Bayesian estimates of the mixture parameters under SRS and RSS samples of size $N=24$ with $H=\{3,4,6\}$ and $\rho=0.9$.}}
 \label{r09_0510}
\end{figure}

Label switching is a common problem in estimating the parameters of the mixture model 
in the Bayesian framework \citep{diebolt1994estimation,stephens2000dealing}. The label 
switching may lead to multimodal posterior distributions of the mixture parameters in 
both SRS and RSS-based Gibbs samplings.   There are various methods in the literature 
for label-switching problems, including, for instance, \cite{celeux2000computational} 
and \cite{stephens2000dealing}. In this paper, following \cite{stephens2000dealing}, 
we imposed the constraint $\mu_1 < \mu_2 < \ldots < \mu_J$ on the MCMC chains followed 
by the post-processing step to mitigate the label switching problem in both SRS and 
RSS methods.   When the MCMC chain remains unimodal, the posterior mode 
was used to estimate the parameter of interest. When posterior distribution appears 
multimodal, we applied the relabeling strategy of \cite{stephens2000dealing} to the 
MCMC chain and relabeled the states to meet the constraint  $\mu_1 < \mu_2 < \ldots < \mu_J$. 
For more information about the relabeling, the readers are referred to  \cite{stephens2000dealing}. 
Because the goal of the manuscript is to compare the performance of the Bayesian method using 
imperfect RSS sampling design with the counterpart under simple random sampling, we have to 
consider the sampling variability  (in the data collection). To do so, we finally replicated 2000 
times the whole data collection and Bayesian estimation procedures.    
  
\begin{figure}
\includegraphics[width=1\textwidth]{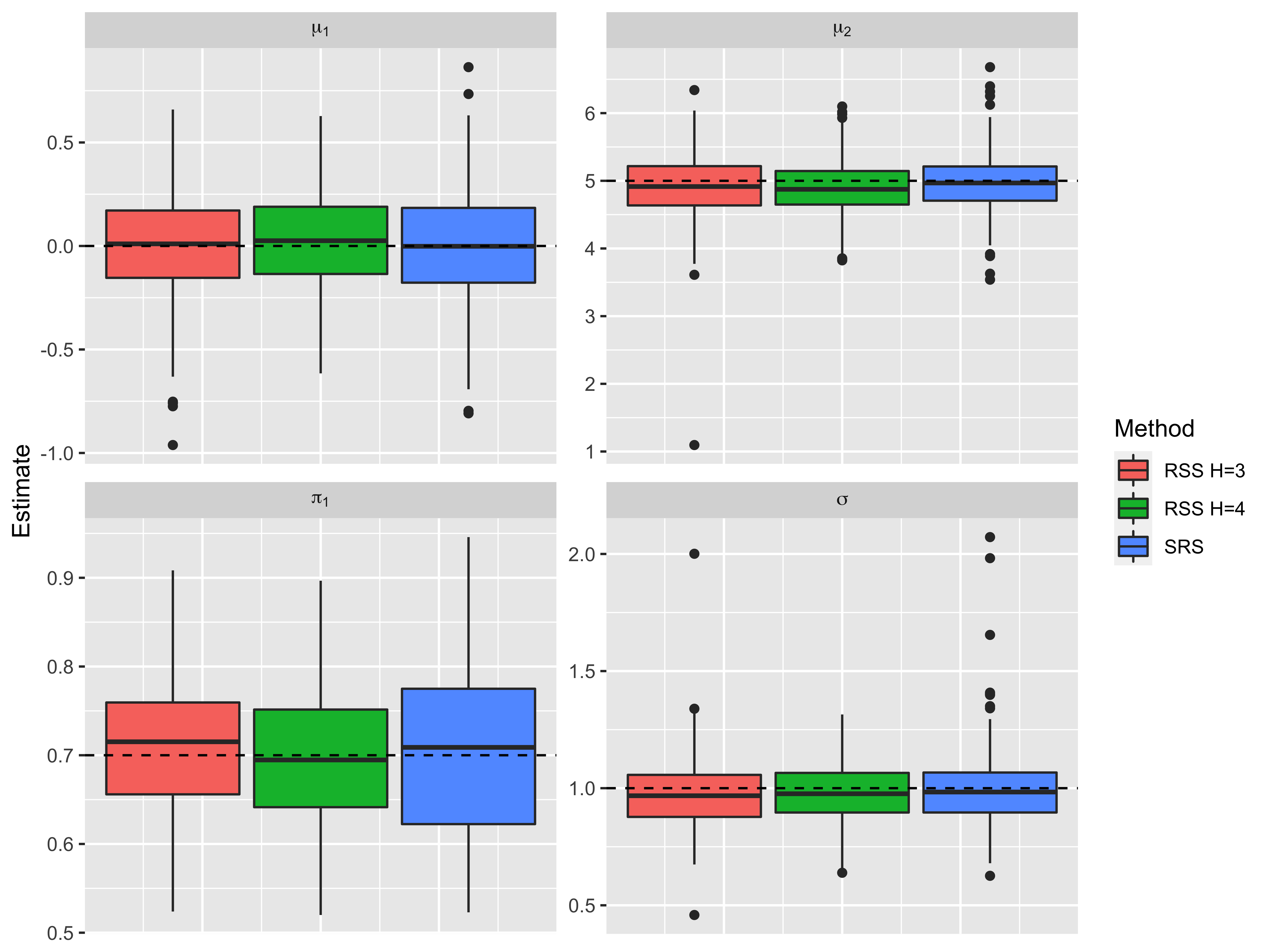}
\caption{\footnotesize{The box-plots for the Bayesian estimates of the mixture parameters under SRS and RSS samples of size $N=24$ with $H=\{3,4\}$ and $\rho=0.7$.}}
 \label{r07_05}
\end{figure}

Figures \ref{r09_05} and \ref{r07_05} show the box-plots of the proposed Bayesian estimation 
methods based on RSS and SRS data of size $N=24$ with $H=\{3,4\}$ for ranking ability $\rho=0.9$ and $\rho=0.7$, respectively.
  We see that the RSS estimators almost always outperform their SRS counterparts. While the 
  centres of the box-plot for both estimation methods are close to the true values of parameters, 
  the variability of  RSS estimates is smaller than that of SRS estimates such that the middle 50\% 
  box of the RSS estimates are almost always contained in that of the SRS estimates. When ranking 
  ability increases,   the variability of the RSS estimators reduces further. Hence,  the  RSS 
  estimators, on average, appear more efficient in estimating the parameters of mixture models.
   
   In addition, we used three other measures to compare the performance of the SRS- and RSS-based 
   Gibbs samplings. These measures include the squared error, the length of the shortest credible 
   interval and the coverage rate of the estimators. We computed the squared distance between the 
   estimate and the true value of the parameter to obtain the squared error. We then reported the 
   $10$, $90$ and $50$ percentiles of the squared errors as the lower (L), upper (U) and middle 
   (M) points for the interval, respectively. To compare the credible intervals of the proposed 
   estimates, we first computed the 95\% shortest credible interval from the posterior distributions. 
   To represent the performance of the credible intervals over 2000 replicates, we computed the 
   95\%  interval for the length of the credible intervals. To do so, we calculated the 2.5, 97.5 
   and 50 percentiles of the length of credible intervals and reported them as the lower (L), upper 
   (U) and middle (M) points for the measure. The coverage rate was measured by the proportion that 
   the shortest credible interval captured the true value of the parameter over 2000 replicates.
  
  Tables \ref{2_comp_rho09} and \ref{2_comp_rho07} show the results of the first simulation study. 
  On average, the RSS-based estimators result in lower squared errors and shorter credible intervals 
  than the SRS-based estimators. The SRS method results in a slightly higher coverage probability 
  than its RSS counterpart, specifically in the estimation of $\sigma^2$. We believe this excellence 
  of SRS is because the prior distributions are conjugate for SRS-based Gibbs sampling. Consequently, 
  the SRS Gibbs sampler enjoys closed marginal posterior distributions for component parameters. 
  Unlike SRS, no closed-form posterior distributions exist for the component parameters in RSS-based 
  Gibbs sampling. In other words, we had to employ an accept-reject step using the Metropolis-Hasting
   approach to sample indirectly from the RSS posterior distributions.   Comparing Tables \ref{2_comp_rho09} 
   and \ref{2_comp_rho07}, one can observe that the squared error, the length of credible intervals 
   and coverage probabilities of the RSS estimators are improved as the ranking ability increases. In 
   addition, when $\rho$ is high,  as the set size increases from $H=3$ to $H=4$, more ranking information 
   is incorporated into RSS data collection. Hence, the Bayesian RSS method's efficiency further improves 
   in estimating the population parameters.     
     
\begin{figure}
\includegraphics[width=1\textwidth]{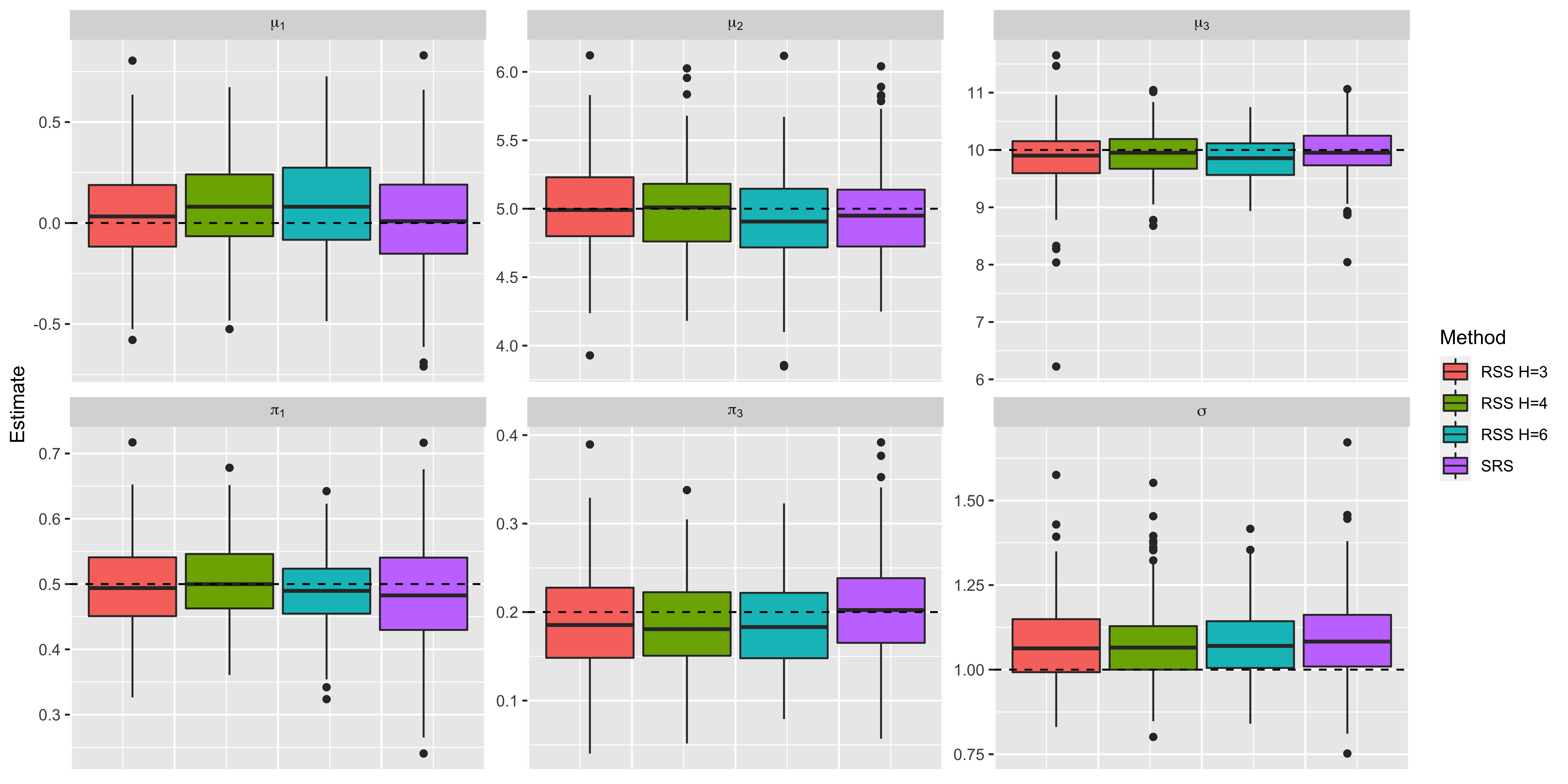}
\caption{\footnotesize{The box-plots for the Bayesian estimates of the mixture parameters under SRS and RSS samples of size $N=24$ with $H=\{3,4,6\}$ and $\rho=0.7$.}}
 \label{r07_0510}
\end{figure}

   In the second simulation study,  we considered the population comprised a mixture of three homoscedastic
    normal distributions with true parameters  $\bpsi_0=(\pi_1,\pi_3,\mu_1,\mu_2,\mu_3,\sigma)=(0.5,0.2,0,5,10,1)$.
     We generated RSS and SRS data of size $N=36$ with set size $H=\{3,4,6\}$ and ranking ability
       $\rho=\{0.7,0.9\}$. As described in the first study, we implemented the two stages of the simulation
        to compute the Bayesian estimates for the parameters of the mixture model using the SRS-based
         Gibbs sampling and RSS-based Metropolis-within-Gibbs sampling. Figures  \ref{r09_0510} and \ref{r07_0510} 
          represent the boxplot of the SRS and RSS posterior modes in estimating the population parameters over
           2000 replicates for $\rho=0.9$ and $\rho=0.7$, respectively. 
            We observe that the box-plot medians are close to the true values of the parameters
            of the population (except for $\sigma^2$) so that the SRS and RSS Bayesian estimates can be considered,
             on average, unbiased in estimating the mixing proportions and the component means. In addition to
              the low bias in both SRS and RSS proposals, we see that RSS-based Gibbs sampling leads to more
               reliable estimates for mixture parameters. Tables \ref{3_comp_rho07} and \ref{3_comp_rho09} represent
                the squared errors, length of the 95\% shortest credible intervals and the coverage probability of
                 the  Bayesian RSS and SRS estimators. The Bayesian RSS method almost always leads to estimates
                  with lower squared errors in estimating population parameters. While the SRS method performs
                   slightly better in the case of coverage rate, particularly in estimating the $\sigma^2$, the
                    coverage probabilities of the two estimation methods are almost close to each other. Last
                     but not least, it is observed that Bayesian RSS estimators provide almost always shorter
                      credible intervals in estimating the mixture parameters.

 \section{Bone Mineral Data Analysis}\label{bone}

As a bone metabolic disease, osteoporosis is recognized by a significant reduction of mass in
 bone tissues. This deterioration of bone architecture results in various major health problems, 
 such as osteoporotic fractures. 
According to the expert panel of WHO, bone mineral density (BMD) is considered one of the 
most reliable risk factors for diagnosing bone disorder status \citep{who2003prevention}. To determine the osteoporosis 
status,  BMD measurements, given by T-score, are compared with the BMD norm of the reference 
group - i.e., the healthy adults between the ages of 20 to 30. 
The status of a patient is diagnosed with osteoporosis if her BMD score is lower than 2.5 
standard deviation from the BMD mean of the reference group \citep{melton1992perspective,burge2007incidence}. While there are plenty of patients 
with osteoporosis, the BMD measurements are obtained from dual-energy X-ray absorptiometry (DXA), 
which requires a time-consuming and expensive procedure. Once acquired, medical experts must 
segment manually and compute the final measurements. While measuring BMD scores is difficult, 
practitioners typically have access to many easy-to-measure patient characteristics, such as 
BMI, age or BMD scores from previous years \citep{cummings1995risk,unnanuntana2010assessment}. According to the cost of BMD measurements, ranked 
set sampling, as a cost-effective sampling technique, can be exploited to incorporate these 
inexpensive characteristics as ranking information into data collection and augment the small 
sample sizes to obtain more efficient estimates for the underlying population.

\begin{figure}
\centering
\includegraphics[width=0.8\textwidth]{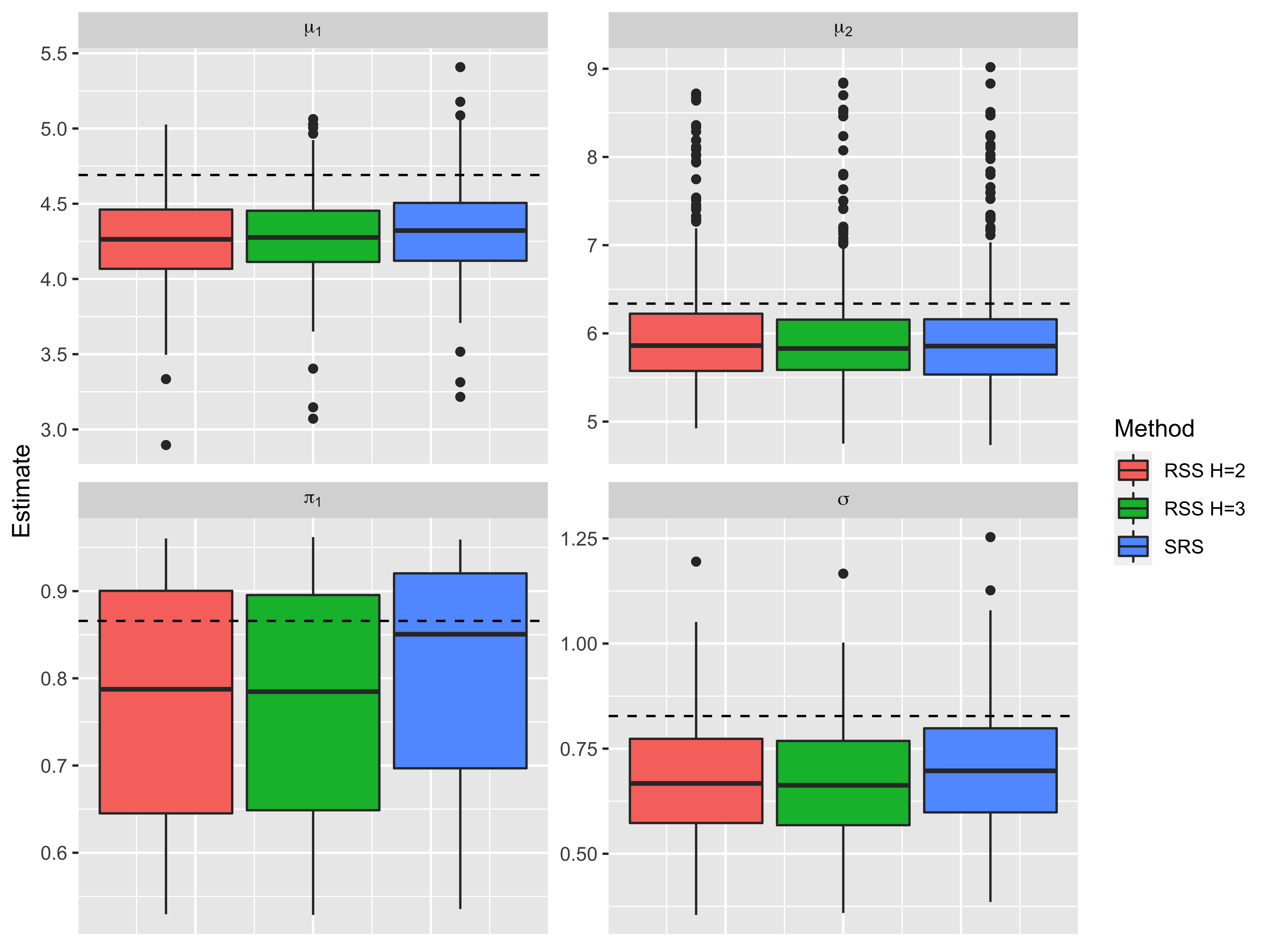}
\caption{\footnotesize{The box-plots for the Bayesian estimates of the BMD population parameters under SRS and RSS samples of size $N=24$ with $H=\{2,3\}$.}}
 \label{fig_real}
\end{figure}

This section applies the Bayesian RSS method to the BMD data from National Health and Nutrition 
Examination Survey (NHANES III). Centres for Disease Control and Prevention (CDC) administered 
the survey in two phases to assess the health and nutritional status of 39695 people between 
1988 to 1994 in the United States. The survey consisted of two bone examinations. There are 182 
women aged 50 and older who participated in both examinations. Owing to the high risk of 
osteoporosis in the aged female population, we treated 182 women as the population in this 
study. We considered the BMD measurement from the femur neck area (FNBMD) from the second 
examination as the outcome $X$. Based on the entire population, the BIC model selection suggests 
a mixture of two normal distributions as the fit to the bone population with parameters 
${\widehat\bpsi}_{B}=(\pi,\mu_1,\mu_2,\sigma)=(0.87,4.69,6.34,0.83)$. We treated the population-level ML estimate ${\widehat\bpsi}_{B}$  as the true parameters 
of the BMD population. 
We used the age of patients as an easy-to-measure concomitant, $Z$, 
with $\rho=cor(X,Z)=-0.49$ to rank the patients in RSS data collection.
 
We applied the Bayesian SRS and RSS methods and estimated the parameters of the BMD population 
using sample size $N=24$ with set size $H=\{2,3\}$ in a similar vein to Section \ref{sim}.
 In the first stage of the numerical study, we used the age of patients to assign judgmental ranks
  involved in RSS data collection and estimated the misplacement probability model $\balpha$. 
  In the second stage, we used the estimated misplacement probability model as the true ranking 
  parameters of RSS to generate RSS data. We also applied the prior distributions \eqref{hom_prior} 
  and then ran the SRS and RSS Gibbs samplers for 15000 iterations with thinning $k=5$ and a burn-in 
  period of $5000$. We eventually replicated the entire data collection and Bayesian estimation procedures 2000 times. 
  
  Figure \ref{fig_real} shows the boxplot of 2000 Bayesian RSS and SRS estimates for the bone population. 
  One observes that the SRS method resulted in a slightly lower bias than the RSS method, particularly in 
estimating the mixing proportion and common variance. We believe this is mainly because the prior 
distributions were conjugate priors for the SRS-based likelihood function. Hence, the SRS Gibbs sampler 
takes advantage of sampling from closed-form posterior distributions for component parameters in each 
iteration. This indirectly also affects the posterior mode of the mixing proportion. Unlike SRS, there 
is no closed for the posterior distributions of component parameters under RSS. The RSS Gibbs sampler 
has to use the Metropolis-Hasting approach to accept or reject the proposals in each state. The 
Bayesian RSS method performs better in estimating the component means and leads to more reliable 
estimates for component parameters of the BMD population. Similar to Section \ref{sim}, we also 
computed the squared error, the length of 95\% shortest 
credible interval and coverage probability for the  Bayesian SRS and RSS estimators. Table 
\ref{table_real} shows the $10$ (L), $50$ (M) and $90$ (U) percentiles of the squared errors, 
2.5 (L), 50 (M) and 97.5 (U) percentiles of the length of 95\% shortest credible intervals as 
well as the coverage rates for the Bayesian estimates of the BMD population. It is observed that 
the SRS Gibbs sampling results in slightly higher coverage rates in estimating the parameters; 
however, the RSS Metropolis within Gibbs sampling almost always leads to lower squared errors 
and shorter credible intervals in estimating the parameters of the BMD population. Therefore, 
when practitioners have access to a decent ranker, the RSS-based Metropolis within Gibbs 
sampling is recommended to estimate the parameters of the bone mineral population.     
    
 \section{Summary and Concluding Remarks}\label{sum}

In many medical surveys, for example, osteoporosis research, measuring the response 
variable (e.g., disease status) is costly, time-consuming or invasive; however, a 
few sampling units can be ranked easily using inexpensive characteristics associated with 
the response variable. In these situations, one can use the ranked set sampling design to 
obtain more representative observations from the population. In this paper, we used RSS 
data to estimate the parameters of the finite mixture of normal distributions in a Bayesian 
framework. Unlike SRS-based Gibbs sampling, there is no closed form for the posterior 
distributions of the component parameters. To cope with this challenge, we developed a 
Metropolis-Hastings approach within Gibbs sampling to draw samples from the conditional 
distributions of mixture parameters under the RSS design.    
In real-life applications, ranking errors are inevitable in RSS sampling. We incorporated 
the ranking errors and ranking information of the RSS data in the Bayesian estimation 
method by a misplacement probability model. 
Since simple random sampling does not require any misplacement ranking model, we treated 
the ranking error probabilities of RSS sampling as fixed and unknown parameters, unlike 
the mixture model parameters. 
Therefore,  We designed an Expectation-Maximization (EM) algorithm within each iteration 
of the Gibbs sampling to estimate the RSS parameters. Through simulation studies, we 
compared the performance of RSS-based Metropolis-within-Gibbs sampling with SRS-based 
Gibbs sampling in estimating the parameters of the mixture of normal distributions. 
Although the SRS method has a slightly higher coverage probability, the RSS method results 
in more reliable estimates with lower squared errors and shorter credible intervals. 
Finally, we applied the developed Bayesian estimators to analyze the bone mineral data of women aged 50 and older.   
  

\nocite{*}
\bibliographystyle{unsrtnat}
\bibliography{Bayes_bib}

\newpage
\section*{Appendix}

\begin{table}[ht]
      \caption{\footnotesize{The $10$ (L), $50$ (M), and $90$ (U) percentiles of squared errors and $2.5$ (L), $50$ (M) and $97.5$ (U) percentiles for length of the 95\% shortest credible interval with coverage probabilities for the Bayesian estimates of the mixture parameters under SRS and RSS samples of size $N=24$ with $H=\{3,4\}$ and $\rho=0.7$.}}
\vspace{0.3cm} 
\centering 
{\footnotesize
\begin{tabular}{c c c c c c c c c c c c c c c}
\hline
\hline
    \multirow{ 2}{*}{Method} &  & \multirow{ 2}{*}{H} & &  \multirow{ 2}{*}{Estimand} & &
    \multicolumn{3}{c}{Squared Error} &  &
    \multicolumn{3}{c}{CI Width} & & \multirow{2}{*}{Coverage} 
    \\
    \cline{7-9}
    \cline{11-13}
     & & & & & & L & M & U & & L & M & U & &\\
    \hline
    \multirow{4}{*}{SRS} & & \multirow{4}{*}{-} & & $\pi_1$ & & 
    0.000 & 0.006 & 0.028 && 0.251 & 0.316 & 0.355 && 0.993 \\  
    & & & & $\mu_1$ & & 
    0.001 & 0.033 & 0.189 && 0.888 & 2.568 & 6.091 && 0.960 \\
    & & & & $\mu_2$ & & 
    0.003 & 0.067 & 0.498 && 1.674 & 5.167 & 6.435 && 0.964 \\  
    & & & & $\sigma$ & & 
    0.000 & 0.008 & 0.053 && 0.491 & 0.629 & 1.138 && 0.964 \\ 
    \hline
    \multirow{4}{*}{RSS} & & \multirow{4}{*}{3} & & $\pi_1$ & & 
    0.000 & 0.003 & 0.019 && 0.251 & 0.289 & 0.320 && 0.974 \\ 
    & & & & $\mu_1$ & & 
    0.001 & 0.028 & 0.151 && 0.712 & 0.985 & 5.598 && 0.940 \\  
    & & & & $\mu_2$ & & 
    0.004 & 0.080 & 0.454 && 1.396 & 1.930 & 6.046 && 0.926 \\ 
    & & & & $\sigma$ & & 
    0.000 & 0.009 & 0.045 && 0.323 & 0.398 & 0.572 && 0.874 \\ 
    \hline
    \multirow{4}{*}{RSS} & & \multirow{4}{*}{4} & & $\pi_1$ & & 
    0.000 & 0.003 & 0.022 && 0.237 & 0.279 & 0.310 && 0.952 \\ 
    & & & & $\mu_1$ & & 
    0.001 & 0.028 & 0.142 && 0.693 & 0.990 & 5.511 && 0.943 \\   
    & & & & $\mu_2$ & & 
    0.002 & 0.069 & 0.432 && 1.339 & 2.079 & 5.996 && 0.933 \\ 
    & & & & $\sigma$ & & 
    0.000 & 0.008 & 0.045 && 0.329 & 0.407 & 0.598 && 0.864 \\
    \hline
    \hline
    \end{tabular}
    }
    \label{2_comp_rho07}
\end{table}

\begin{table}[ht]
\caption{\footnotesize{The $10$ (L), $50$ (M), and $90$ (U) percentiles of squared errors and $2.5$ (L), $50$ (M) and $97.5$ (U) percentiles for length of the 95\% shortest credible interval with coverage probabilities for the Bayesian estimates of the BMD population parameters under SRS and RSS samples of size $N=24$ with $H=\{2,3\}$.}}
\vspace{0.3cm} 
\centering 
{\footnotesize
\begin{tabular}{c c c c c c c c c c c c c c c}
\hline
\hline
    \multirow{ 2}{*}{Method} &  & \multirow{ 2}{*}{H} & &  \multirow{ 2}{*}{Estimand} & &
    \multicolumn{3}{c}{Squared Error} &  &
    \multicolumn{3}{c}{CI Width} & & \multirow{2}{*}{Coverage} 
    \\
    \cline{7-9}
    \cline{11-13}
     & & & & & & L & M & U & & L & M & U & &\\
    \hline
    \multirow{4}{*}{SRS} & & \multirow{4}{*}{-} & & $\pi_1$ & & 
    0.000 & 0.006 & 0.097 && 0.376 & 0.467 & 0.477 && 0.996 \\   
    & & & & $\mu_1$ & & 
    0.003 & 0.189 & 0.828 && 0.852 & 1.531 & 2.147 && 0.985 \\ 
    & & & & $\mu_2$ & & 
    0.011 & 0.358 & 1.780 && 2.116 & 3.507 & 6.186 && 0.803 \\   
    & & & & $\sigma$ & & 
    0.000 & 0.021 & 0.112 && 0.437 & 0.611 & 0.787 && 0.949 \\
    \hline
    \multirow{4}{*}{RSS} & & \multirow{4}{*}{2} & & $\pi_1$ & & 
    0.000 & 0.008 & 0.103 && 0.307 & 0.459 & 0.475 && 0.947 \\  
    & & & & $\mu_1$ & & 
    0.004 & 0.183 & 0.719 && 0.681 & 1.521 & 2.216 && 0.979 \\  
    & & & & $\mu_2$ & & 
    0.003 & 0.336 & 1.674 && 1.970 & 3.233 & 5.281 && 0.748 \\ 
    & & & & $\sigma$ & & 
    0.000 & 0.027 & 0.135 && 0.300 & 0.468 & 0.623 && 0.839 \\ 
    \hline
    \multirow{4}{*}{RSS} & & \multirow{4}{*}{3} & & $\pi_1$ & & 
    0.000 & 0.008 & 0.101 && 0.318 & 0.460 & 0.475 && 0.943 \\ 
    & & & & $\mu_1$ & & 
    0.003 & 0.172 & 0.710 && 0.685 & 1.448 & 2.192 && 0.979 \\   
    & & & & $\mu_2$ & & 
    0.004 & 0.320 & 1.369 && 2.002 & 3.200 & 5.111 && 0.756 \\ 
    & & & & $\sigma$ & & 
    0.000 & 0.028 & 0.133 && 0.323 & 0.457 & 0.599 && 0.803 \\
    \hline
    \hline
    \end{tabular}
    }
    \label{table_real}
\end{table}

\begin{table}[ht]
      \caption{\footnotesize{The $10$ (L), $50$ (M), and $90$ (U) percentiles of squared errors and $2.5$ (L), $50$ (M) and $97.5$ (U) percentiles for length of the 95\% shortest credible interval with coverage probabilities for the Bayesian estimates of the mixture parameters under SRS and RSS samples of size $N=36$ with $H=\{3,4,6\}$ and $\rho=0.9$.}}
\vspace{0.3cm} 
\centering 
{\footnotesize
\begin{tabular}{c c c c c c c c c c c c c c c}
\hline
\hline
    \multirow{ 2}{*}{Method} &  & \multirow{ 2}{*}{H} & &  \multirow{ 2}{*}{Estimand} & &
    \multicolumn{3}{c}{Squared Error} &  &
    \multicolumn{3}{c}{CI Width} & & \multirow{2}{*}{Coverage} 
    \\
    \cline{7-9}
    \cline{11-13}
     & & & & & & L & M & U & & L & M & U & &\\
    \hline
    \multirow{6}{*}{SRS} & & \multirow{6}{*}{-}& & $\pi_1$ 
    & & 0.000 & 0.003 & 0.022 && 0.301 & 0.317 & 0.492 && 0.960 \\ 
    & & & & $\pi_3$ & & 0.000 & 0.001 & 0.010 && 0.229 & 0.264 & 0.338 && 0.980 \\
    & & & & $\mu_1$ & & 0.001 & 0.033 & 0.188 && 0.914 & 1.167 & 2.395 && 0.980 \\ 
    & & & & $\mu_2$ & & 0.001 & 0.044 & 0.306 && 1.164 & 1.692 & 5.459 && 0.975 \\
    & & & & $\mu_3$ & & 0.003 & 0.070 & 0.396 && 1.418 & 1.855 & 3.669 && 0.985 \\
    & & & & $\sigma$ & & 0.000 & 0.008 & 0.063 && 0.482 & 0.637 & 1.461 && 0.854 \\
    \hline
    \multirow{6}{*}{RSS} & & \multirow{6}{*}{3}& & $\pi_1$ 
    & & 0.000 & 0.003 & 0.013 && 0.251 & 0.266 & 0.308 && 0.960 \\ 
    & & & & $\pi_3$ & & 0.000 & 0.001 & 0.008 && 0.189 & 0.222 & 0.258 && 0.975 \\
    & & & & $\mu_1$ & & 0.001 & 0.032 & 0.192 && 0.807 & 1.009 & 1.391 && 0.945 \\
    & & & & $\mu_2$ & & 0.001 & 0.054 & 0.286 && 1.068 & 1.504 & 2.980 && 0.980 \\ 
    & & & & $\mu_3$ & & 0.001 & 0.067 & 0.444 && 1.378 & 1.766 & 2.706 && 0.975 \\
    & & & & $\sigma$ & & 0.000 & 0.008 & 0.046 && 0.324 & 0.412 & 0.869 && 0.794 \\
    \hline
    \multirow{6}{*}{RSS} & & \multirow{6}{*}{4}& & $\pi_1$ 
    & & 0.000 & 0.002 & 0.012 && 0.238 & 0.254 & 0.293 && 0.965 \\
    & & & & $\pi_3$ & & 0.000 & 0.001 & 0.009 && 0.184 & 0.213 & 0.265 && 0.960 \\
    & & & & $\mu_1$ & & 0.001 & 0.024 & 0.126 && 0.799 & 0.990 & 1.336 && 0.955 \\
    & & & & $\mu_2$ & & 0.001 & 0.045 & 0.267 && 1.049 & 1.466 & 3.408 && 0.980 \\  
    & & & & $\mu_3$ & & 0.003 & 0.074 & 0.586 && 1.350 & 1.738 & 2.888 && 0.965 \\
    & & & & $\sigma$ & & 0.000 & 0.009 & 0.051 && 0.324 & 0.428 & 0.856 && 0.754 \\
    \hline
    \multirow{6}{*}{RSS} & & \multirow{6}{*}{6}& & $\pi_1$ 
    & & 0.000 & 0.001 & 0.011 && 0.217 & 0.233 & 0.255 && 0.965 \\
    & & & & $\pi_3$ & & 0.000 & 0.001 & 0.007 && 0.169 & 0.193 & 0.220 && 0.935 \\
    & & & & $\mu_1$ & & 0.001 & 0.021 & 0.133 && 0.732 & 0.873 & 1.092 && 0.955 \\
    & & & & $\mu_2$ & & 0.002 & 0.050 & 0.301 && 1.041 & 1.321 & 2.199 && 0.980 \\  
    & & & & $\mu_3$ & & 0.002 & 0.072 & 0.504 && 1.211 & 1.614 & 2.270 && 0.970 \\ 
    & & & & $\sigma$ & & 0.000 & 0.005 & 0.033 && 0.307 & 0.390 & 0.647 && 0.854 \\
    \hline
    \hline
    \end{tabular}
    }
    \label{3_comp_rho09}
\end{table}

\begin{table}[ht]
      \caption{\footnotesize{The $10$ (L), $50$ (M), and $90$ (U) percentiles of squared errors and $2.5$ (L), $50$ (M) and $97.5$ (U) percentiles for length of the 95\% shortest credible interval with coverage probabilities for the Bayesian estimates of the mixture parameters under SRS and RSS samples of size $N=36$ with $H=\{3,4,6\}$ and $\rho=0.7$.}}
      \vspace{0.3cm} 
\centering 
{\footnotesize
\begin{tabular}{c c c c c c c c c c c c c c c}
\hline
\hline
    \multirow{ 2}{*}{Method} &  & \multirow{ 2}{*}{H} & &  \multirow{ 2}{*}{Estimand} & &
    \multicolumn{3}{c}{Squared Error} &  &
    \multicolumn{3}{c}{CI Width} & & \multirow{2}{*}{Coverage} 
    \\
    \cline{7-9}
    \cline{11-13}
     & & & & & & L & M & U & & L & M & U & &\\
    \hline
    \multirow{6}{*}{SRS} & & \multirow{6}{*}{-}& & $\pi_1$ 
    & & 0.000 & 0.003 & 0.022 && 0.301 & 0.317 & 0.492 && 0.960 \\ 
    & & & & $\pi_3$ & & 0.000 & 0.001 & 0.010 && 0.229 & 0.264 & 0.338 && 0.980 \\
    & & & & $\mu_1$ & & 0.001 & 0.033 & 0.188 && 0.914 & 1.167 & 2.395 && 0.980 \\ 
    & & & & $\mu_2$ & & 0.001 & 0.044 & 0.306 && 1.164 & 1.692 & 5.459 && 0.975 \\
    & & & & $\mu_3$ & & 0.003 & 0.070 & 0.396 && 1.418 & 1.855 & 3.669 && 0.985 \\
    & & & & $\sigma$ & & 0.000 & 0.008 & 0.063 && 0.482 & 0.637 & 1.461 && 0.854 \\
    \hline
    \multirow{6}{*}{RSS} & & \multirow{6}{*}{3}& & $\pi_1$ 
    & & 0.000 & 0.002 & 0.010 && 0.251 & 0.268 & 0.334 && 0.989 \\ 
    & & & & $\pi_3$ & & 0.000 & 0.002 & 0.009 && 0.187 & 0.221 & 0.278 && 0.967 \\
    & & & & $\mu_1$ & & 0.001 & 0.028 & 0.187 && 0.807 & 0.998 & 1.435 && 0.960 \\
    & & & & $\mu_2$ & & 0.002 & 0.042 & 0.354 && 1.100 & 1.480 & 3.954 && 0.989 \\  
    & & & & $\mu_3$ & & 0.002 & 0.080 & 0.611 && 1.394 & 1.745 & 2.980 && 0.971 \\
    & & & & $\sigma$ & & 0.000 & 0.007 & 0.050 && 0.322 & 0.421 & 0.841 && 0.772 \\
    \hline
    \multirow{6}{*}{RSS} & & \multirow{6}{*}{4}& & $\pi_1$ 
    & & 0.000 & 0.002 & 0.010 && 0.240 & 0.256 & 0.301 && 0.971 \\
    & & & & $\pi_3$ & & 0.000 & 0.002 & 0.008 && 0.181 & 0.209 & 0.257 && 0.949 \\
    & & & & $\mu_1$ & & 0.001 & 0.027 & 0.149 && 0.785 & 0.951 & 1.358 && 0.953 \\
    & & & & $\mu_2$ & & 0.001 & 0.051 & 0.325 && 1.057 & 1.472 & 3.158 && 0.978 \\  
    & & & & $\mu_3$ & & 0.003 & 0.069 & 0.499 && 1.333 & 1.784 & 2.735 && 0.960 \\
    & & & & $\sigma$ & & 0.000 & 0.007 & 0.051 && 0.324 & 0.409 & 0.875 && 0.801 \\
    \hline
    \multirow{6}{*}{RSS} & & \multirow{6}{*}{6}& & $\pi_1$ 
    & & 0.000 & 0.001 & 0.009 && 0.222 & 0.240 & 0.303 && 0.960 \\
    & & & & $\pi_3$ & & 0.000 & 0.001 & 0.007 && 0.174 & 0.197 & 0.238 && 0.980 \\
    & & & & $\mu_1$ & & 0.001 & 0.035 & 0.189 && 0.776 & 0.934 & 1.358 && 0.935 \\
    & & & & $\mu_2$ & & 0.002 & 0.065 & 0.301 && 1.099 & 1.445 & 3.029 && 0.965 \\  
    & & & & $\mu_3$ & & 0.001 & 0.082 & 0.516 && 1.325 & 1.682 & 2.847 && 0.960 \\
    & & & & $\sigma$ & & 0.000 & 0.007 & 0.051 && 0.330 & 0.413 & 0.903 && 0.794 \\
    \hline
    \hline
    \end{tabular}
    }
    \label{3_comp_rho07}
\end{table}

\begin{table}[ht]
      \caption{\footnotesize{The true values, bias and MSE for the ML estimates of misplacement probabilities based on the RSS data of size $N=36$ with set size $H=3$ and ranking ability $\rho=0.7,0.9$.}}
\vspace{0.3cm} 
\centering 
{\footnotesize
\begin{tabular}{c c c c c c c c c c}
\hline
\hline
    $\rho$ & & Estimated & &  True value & & Absolute bias & &
    MSE 
    \\
    \hline
    \multirow{3}{*}{0.7} & & $\alpha_{11}$ & & 0.8300 & & 0.1180 & & 0.0176 \\
     & & $\alpha_{21}$ & & 0.1536 & & 0.1126 & & 0.0159 \\
     & & $\alpha_{22}$ & & 0.7416 & & 0.1615 & & 0.0340 \\
    \hline
    \multirow{3}{*}{0.9} & & $\alpha_{11}$ & & 0.9041 & & 0.0854 & & 0.0104 \\
     & & $\alpha_{21}$ & & 0.0902 & & 0.0854 & & 0.0103 \\
     & & $\alpha_{22}$ & & 0.8483 & & 0.1178 & & 0.0212 \\
    \hline
    \hline
    \end{tabular}
    }
    \label{alpha}
\end{table}

\end{document}